\documentclass{pasj01}
\usepackage{longtable}
\usepackage{graphicx} 
\usepackage{natbib} 
\usepackage{enumerate}
\usepackage{xspace}
\usepackage{xcolor}
\usepackage{comment}
\usepackage[left]{lineno}

\newcommand{\target}{Ross\,508\xspace}

\newcommand{\mear}{M_{\oplus}}
\newcommand{\mps}{{\rm m\,s}^{-1}}

\Received{$\langle$March 5, 2022$\rangle$}
\Accepted{$\langle$May 23, 2022$\rangle$}
\Published{$\langle$publication date$\rangle$}

\begin{document}

\title{A Super-Earth Orbiting Near the Inner Edge of the Habitable Zone around the M4.5-dwarf \target}
\author{Hiroki \textsc{Harakawa} \altaffilmark{1}
Takuya \textsc{Takarada} \altaffilmark{2,3}
Yui \textsc{Kasagi} \altaffilmark{4}
Teruyuki \textsc{Hirano} \altaffilmark{2,3,4}
Takayuki \textsc{Kotani} \altaffilmark{2,3,4}
Masayuki \textsc{Kuzuhara} \altaffilmark{2,3}
Masashi \textsc{Omiya} \altaffilmark{2,3}
Hajime \textsc{Kawahara} \altaffilmark{5}
Akihiko \textsc{Fukui} \altaffilmark{6}
Yasunori \textsc{Hori} \altaffilmark{2,3,4}
Hiroyuki Tako \textsc{Ishikawa} \altaffilmark{2,3}
Masahiro \textsc{Ogihara} \altaffilmark{7,8,3}
John \textsc{Livingston} \altaffilmark{2,3,9}
Timothy D. \textsc{Brandt} \altaffilmark{10}
Thayne \textsc{Currie} \altaffilmark{1,11,12}
Wako \textsc{Aoki} \altaffilmark{3}
Charles A. \textsc{Beichman} \altaffilmark{13,14}
Thomas \textsc{Henning} \altaffilmark{15}
Klaus \textsc{Hodapp} \altaffilmark{16}
Masato \textsc{Ishizuka} \altaffilmark{9}
Hideyuki \textsc{Izumiura} \altaffilmark{17}
Shane \textsc{Jacobson} \altaffilmark{16}
Markus \textsc{Janson} \altaffilmark{18}
Eiji \textsc{Kambe} \altaffilmark{1}
Takanori \textsc{Kodama} \altaffilmark{6}
Eiichiro \textsc{Kokubo} \altaffilmark{3}
Mihoko \textsc{Konishi} \altaffilmark{19}
Vigneshwaran \textsc{Krishnamurthy} \altaffilmark{2,3}
Tomoyuki \textsc{Kudo} \altaffilmark{1}
Takashi \textsc{Kurokawa} \altaffilmark{2,20}
Nobuhiko \textsc{Kusakabe} \altaffilmark{2,3}
Jungmi \textsc{Kwon} \altaffilmark{9}
Yuji \textsc{Matsumoto} \altaffilmark{3}
Michael W. \textsc{McElwain} \altaffilmark{21}
Koyu \textsc{Mitsui} \altaffilmark{9}
Takao \textsc{Nakagawa} \altaffilmark{22}
Norio \textsc{Narita} \altaffilmark{2,6,23}
Jun \textsc{Nishikawa} \altaffilmark{3,4,2}
Stevanus K. \textsc{Nugroho} \altaffilmark{2,3}
Eugene \textsc{Serabyn} \altaffilmark{24}
Takuma \textsc{Serizawa} \altaffilmark{20, 3}
Aoi \textsc{Takahashi} \altaffilmark{2,3}
Akitoshi \textsc{Ueda} \altaffilmark{2,3,4}
Taichi \textsc{Uyama} \altaffilmark{13,14,3}
S\'{e}bastien \textsc{Vievard} \altaffilmark{1,2}
Ji \textsc{Wang} \altaffilmark{25}
John \textsc{Wisniewski} \altaffilmark{26}
Motohide \textsc{Tamura} \altaffilmark{9,2,3}
Bun'ei \textsc{Sato} \altaffilmark{27}
}%

\altaffiltext{1}{Subaru Telescope, 650 North A'ohoku Place, Hilo HI 96720, USA}
\altaffiltext{2}{Astrobiology Center, 2-21-1 Osawa, Mitaka, Tokyo 181-8588, Japan}
\altaffiltext{3}{National Astronomical Observatory of Japan, 2-21-1 Osawa, Mitaka, Tokyo 181-8588, Japan}
\altaffiltext{4}{Department of Astronomical Science, The Graduate University for Advanced Studies, SOKENDAI, 2-21-1 Osawa, Mitaka, Tokyo 181-8588, Japan}
\altaffiltext{5}{Department of Earth and Planetary Sciences, Graduate School of Science, The University of Tokyo, 7-3-1 Hongo, Bunkyo-ku, Tokyo 113-0033, Japan}
\altaffiltext{6}{Komaba Institute for Science, The University of Tokyo, 3-8-1 Komaba, Meguro, Tokyo 153-8902, Japan}
\altaffiltext{7}{Tsung-Dao Lee Institute, Shanghai Jiao Tong University, 520 Shengrong Road, Shanghai 201210, China}
\altaffiltext{8}{Earth-Life Science Institute, Tokyo Institute of Technology, 2-12-1 Ookayama, Meguro, Tokyo 152-8550, Japan}
\altaffiltext{9}{Department of Astronomy, Graduate School of Science, The University of Tokyo, 7-3-1 Hongo, Bunkyo-ku, Tokyo 113-0033, Japan}
\altaffiltext{10}{Department of Physics, University of California, Santa Barbara, Santa Barbara, CA 93106, USA}
\altaffiltext{11}{NASA-Ames Research Center, Moffett Field, CA, USA}
\altaffiltext{12}{Eureka Scientific, Oakland, 10 CA, USA}
\altaffiltext{13}{Infrared Processing and Analysis Center, California Institute of Technology, 1200 E. California Boulevard, Pasadena, CA 91125, USA}
\altaffiltext{14}{NASA Exoplanet Science Institute, Pasadena, CA 91125, USA}
\altaffiltext{15}{Max-Planck-Institut f\"{u}r Astronomie, K\"{o}nigstuhl 17, 69117 Heidelberg, Germany}
\altaffiltext{16}{University of Hawaii, Institute for Astronomy, 640 N. Aohoku Place, Hilo, HI 96720, USA}
\altaffiltext{17}{Okayama Branch, Subaru Telescope, National Astronomical Observatory of Japan, National Institute of Natural Sciences, Kamogata, Asakuchi, Okayama 719-0232, Japan}
\altaffiltext{18}{Department of Astronomy, Stockholm University, SE-10691, Stockholm, Sweden}
\altaffiltext{19}{Faculty of Science and Technology, Oita University, 700 Dannoharu, Oita 870-1192, Japan}
\altaffiltext{20}{Institute of Engineering, Tokyo University of Agriculture and Technology, 2-24-16, Nakacho, Koganei, Tokyo, 184-8588, Japan}
\altaffiltext{21}{Exoplanets and Stellar Astrophysics Laboratory, NASA Goddard Space Flight Center, Greenbelt, MD 20771, USA}
\altaffiltext{22}{Institute of Space and Astronautical Science, Japan Aerospace Exploration Agency, 3-1-1 Yoshinodai, Chuo-ku, Sagamihara, Kanagawa 252-5210, Japan}
\altaffiltext{23}{Instituto de Astrof\'{i}sica de Canarias (IAC), 38205 La Laguna, Tenerife, Spain}
\altaffiltext{24}{Jet Propulsion Laboratory, California Institute of Technology, Pasadena, California 91109, USA}
\altaffiltext{25}{Department of Astronomy, The Ohio State University, 100 W 18th Ave, Columbus, OH 43210 USA}
\altaffiltext{26}{Department of Physics and Astronomy, University of Oklahoma, 440 West Brooks Street, Norman, OK 73019, USA}
\altaffiltext{27}{Department of Earth and Planetary Sciences, Tokyo Institute of Technology, 2-12-1 Ookayama, Meguro-ku, Tokyo 152-8551, Japan}

\email{harakawa@naoj.org}


\maketitle

\begin{abstract}

We report the near-infrared radial-velocity (RV) discovery of a super-Earth planet on a 10.77-day orbit around the M4.5 dwarf \target ($J_\mathrm{mag}=9.1$). Using precision RVs from the Subaru Telescope IRD (InfraRed Doppler) instrument, we derive a semi-amplitude of $3.92^{+0.60}_{-0.58}$ $\mps$, corresponding to a planet with a minimum mass $m \sin i = 4.00^{+0.53}_{-0.55}\ \mear$. We find no evidence of significant signals at the detected period in spectroscopic stellar activity indicators or MEarth photometry. The planet, \target~b, has a semimajor-axis of $0.05366^{+0.00056}_{-0.00049}$ au.
This gives an orbit-averaged insolation of $\approx$1.4 times the Earth's value, placing \target b near the inner edge of its star's habitable zone.
We have explored the possibility that the planet has a high eccentricity and its host is accompanied by an additional unconfirmed companion on a wide orbit.
Our discovery demonstrates that the near-infrared RV search can play a crucial role to find a low-mass planet around cool M dwarfs like Ross 508.
\end{abstract}

\section{Introduction}
Since the discovery of 51 Pegasi b around a solar-type star \citep{1995Natur.378..355M}, precision radial velocity (RV) searches have discovered nearly a thousand exoplanets \citep{Schneider_2011_exoplaneteu}. 
More recently, transit surveys, with observatories including 
\textit{CoRoT} \citep{2006ESASP1306...33B}, \textit{Kepler} \citep{2010Sci...327..977B}, and \textit{TESS} \citep{2015JATIS...1a4003R} have discovered several thousand more. Exoplanets are known to orbit various types of stars such as solar-type stars \citep[e.g.][]{2011Natur.470...53L}, low-mass M dwarfs \citep{2017Natur.542..456G}, evolved stars \citep{2022PASJ...74...92T}, and stellar remnants \citep[e.g.][]{2020Natur.585..363V}. 
Among them, M-type stars are especially promising targets for the detection of Earth-like planets.
These stars' small sizes make transits relatively deep, and their low luminosities make the habitable zone close to the star where the RV amplitude is larger. 
\par

Nevertheless, exoplanet discoveries around cool M dwarfs are still limited.\footnote{
Only 3 (2) stars with effective temperatures less than 3000 K have been discovered to host planets via the RV (transit) technique, according to a query of the NASA Exoplanet Archive in February 2022.
Note that the effective temperatures from the TESS Input Catalog \citep[TIC;][]{Stassun_2019} were adopted for the majority of the sample.}
Most exoplanet surveys have used optical CCDs in their cameras but such cool stars emit most of their energy in the near-infrared (NIR).
One of the most effective ways to search for planets around cool M-type stars is to use an infrared-sensitive high-dispersion and high-precision spectrograph. 
Recently, several teams have commissioned NIR spectrographs for high-precision RV surveys, including CARMENES \citep[Calar Alto high-Resolution search for M dwarfs with Exoearths with Near-infrared and optical Echelle Spectrographs;][]{2016SPIE.9908E..12Q}, HPF \citep[Habitable Planet Finder;][]{2014SPIE.9147E..1GM}, and SPIROU \citep[SPectropolarimetre InfraROUge;][]{2012SPIE.8446E..30T}. 
The RV surveys performed with those spectrographs have so far reported a few detections of planetary systems around M dwarfs cooler than $\sim$3000 K \citep[e.g., ][]{Zechmeister_2019_Teegarden}, while they have reported dozens of exoplanets around stars with effective temperature higher than $\sim$3000 K.
It is notable that optical RV measurements have been primarily used for those detections; for example, the terrestrial planets around Teegarden's star were discovered using the optical channel of CARMENES \citep[][]{Zechmeister_2019_Teegarden}. 
High-precision RV measurements in the NIR facilitate the detection of planets around cooler M dwarfs, which remains a frontier in exoplanet exploration. \par

IRD (InfraRed Doppler instrument) is a high-precision, high-dispersion ($R = 70,000$) NIR spectrograph mounted on the Subaru 8.2-m telescope \citep{2012SPIE.8446E..1TT,2018SPIE10702E..11K}. 
To achieve a velocity precision of 2--3 $\mps$, IRD is aided by a wide-band laser-frequency comb \citep[LFC: ][]{Kashiwagi:16,Kokubo:2016}, and an adaptive optics, enabling the use of a narrow slit-width.
The combination of a large-aperture telescope with high RV precision in the NIR thus makes IRD one of the best instruments for studying cool stars, in particular late M dwarfs, whose flux peaks are located in the NIR. 
In February 2019, we started an extensive RV survey program for nearby mid-to-late M-type dwarfs within the Subaru Strategic Program (SSP; \citealt{2018SSPproposal..S}) framework.
This program employs IRD with the aim of detecting planets down to Earth-mass in the habitable zones (HZs) of nearby late M dwarfs. 
The capabilities of IRD allow the systematic survey of fainter, and thus later-type, M dwarfs than ever before. \par

In this paper, we present the first exoplanet discovery from the IRD-SSP campaign, a super Earth that orbits near the inner edge of the HZ around \target\ (the star is also known as LSPM J1523+1727), which is an M4.5-type dwarf \citep{Koizumi_2018_Mdwarf} at a distance of 11.2 pc \citep{Gaia_2021_eDR3} from Earth. 
In Section \ref{Obs}, we describe the observations and data reduction of \target. In Section \ref{Ana_Res} we present our analysis of the fundamental properties and activity of \target, along with the determination of the planet's orbit from the RV measurements. 
Finally, in Section \ref{Discussion}, we discuss the uniqueness of the planet and its potential formation processes, concluding with a summary.

\section{Observations}
\label{Obs}

\subsection{Target Selection}
\target was observed as part of the IRD-SSP survey because of its low mass ($M<0.25~M_\odot$), low temperature ($T_{\rm eff} < 3400$~K), low $v\sin i$ ($<$5~km\,s$^{-1}$), and low stellar activity.  
The initial target list was prepared based on literature measurements satisfying the above criteria \citep{2018SSPproposal..S} supplemented with optical medium-resolution spectroscopic observations \citep{Koizumi_2018_Mdwarf}.
Stars with no rotation period and $v\sin{i}$ measurements were required to have nondetections of H$\alpha$ emission, which is expected for inactive and slowly-rotating stars.
We continually refine our target list, dropping stars from our long-term monitoring campaign if IRD spectra show them to be double-lined spectroscopic binaries or rapid rotators, or if we detect large RV variations suggestive of stellar companions. 
With these screening data, we plan to select about 60 mid and late M dwarfs with low RV variability and high RV precision for the RV monitors, after about three-year observations of its planned five-year survey period.

\subsection{Observations and Data Reduction}
\label{obs_and_red}
We obtained 102 high-resolution, high S/N spectra of \target\ using IRD over $\approx$3 years from 2019 to 2021. 
All stellar spectra were obtained simultaneously with LFC spectra to provide a fiducial wavelength reference for precision RV measurements. 
The typical exposure time for each frame was 600 seconds, achieving an S/N ratio of about 90 per pixel at $1~\mathrm{\mu m}$ wavelength.

The 2 H2RG (HAWAII-2RG) detectors installed in IRD show mutually independent bias levels for each readout channel. 
We thus used our bias subtraction code optimized for those two detectors to suppress bias counts \citep{2018SPIE10702E..60K}. 
We also subtracted correlated read noise by applying a commonly used technique for H2RG detectors \citep[e.g.,][]{Brandt_2013_ACORNS} to the science pixels in our images with the temporal masks to the 2D-spectra. 

Following the removal of bias and read noise, we used \texttt{IRAF} (\texttt{echelle} package) for subsequent \'echelle data reduction procedures, such as scattered light subtraction, flat fielding, and extraction of one-dimensional spectra.
Preliminary wavelength calibrations were done using Th-Ar spectra, but we obtained precise RV measurements using LFC spectra \citep[see][for details]{2020PASJ...72...93H}.
Details of the RV measurements from the 1D spectra are described in Section \ref{rv_mesurements}.

\section{Analysis and Results}\label{Ana_Res}

\subsection{Stellar Parameters}

We derive the fundamental stellar parameters for \target\ using a combination of literature measurements and IRD spectra.  
Table \ref{tab:stellarparam} summarizes all of our adopted stellar parameters including the ones we derive below.

\begin{table}
  \tbl{Stellar parameters of \target}{
    \begin{tabular}{lcc}
    \hline\noalign{\vskip2.3pt}
        Parameter & Value & References  \\[2pt]
    \hline\noalign{\vskip3pt}
    $\alpha$ (J2000.0)  &  \timeform{15h23m50.699s}  &  \textit{Gaia} eDR3  \\
    $\delta$ (J2000.0)  &  \timeform{+17D27'37.30''}  &  \textit{Gaia} eDR3  \\
    $\varpi$ (mas)  &  $89.1284\pm0.0331$  &  \textit{Gaia} eDR3  \\
    Distance (pc) & $11.2183 \pm 0.0035$ & \citet{2021AJ....161..147B}\\
    RUWE & 1.487  &  \textit{Gaia} eDR3  \\
    $G$ (mag)  &  $12.1952\pm0.0029$  &  \textit{Gaia} eDR3  \\
    $G_\mathrm{BP}$ (mag)  &  $13.9882\pm0.0044$  &  \textit{Gaia} eDR3  \\
    $G_\mathrm{RP}$ (mag)  &  $10.9204\pm0.0042$  &  \textit{Gaia} eDR3  \\
    $J$ (mag)  &  $9.105\pm0.024$  &  2MASS  \\
    $K_\mathrm{s}$ (mag)  &  $8.279\pm0.023$  &  2MASS  \\
    Spectral Type &  M4.5  &  \cite{Koizumi_2018_Mdwarf}  \\
    $T_\mathrm{eff}$ (K)  &  $3071^{+34}_{-22}$  &  This work \\
    $\log g$ (cgs)  &  $5.039\pm 0.027$  &  This work \\
    $L_\star$ ($L_\odot$)  &  $3.589^{+0.067}_{-0.071} \times 10^{-3}$  &  This work \\
    $M_\star$ ($M_\odot$)  &  $0.1774 \pm 0.0045$  &  This work \\
    $R_\star$ ($R_\odot$)  &  $0.2113 \pm 0.0063$  &  This work \\
    $\rho_\star$ (g cm$^{-3}$)  &  $26.5_{-2.3}^{+2.5}$  &  This work  \\  
    $\mathrm{[Fe/H]}$ (dex)  &  $-0.20\pm0.20$  &  \citet{2022AJ....163...72I} \\ 
    \hline\noalign{\vskip0pt}
  \end{tabular}}
  \label{tab:stellarparam}
  \begin{tabnote}
  \end{tabnote}
\end{table}

For \target's metallicity, we adopt its iron abundance [Fe/H] determined by \citet{2022AJ....163...72I} from the same IRD spectra that we use here. 
They conducted the equivalent width analysis on the atomic absorption lines of Na, Mg, Ca, Ti, Cr, Mn, Fe, and Sr to derive individual elemental abundances that are consistent with each other.
The abundance of individual elements will help to constrain the detailed geophysical properties of the planets, although it is beyond the scope of this paper.
\target is a relatively metal-poor star, but its abundance ratio of each element is consistent with the solar composition within the errors.
Their abundance and kinematic analyses show characteristics between the Galactic thin and thick disks, suggesting the possibility of a relatively old population. \par

We next analyzed the spectral energy distribution (SED) of \target\ to estimate its effective temperature and luminosity. 
The SED was calculated from the magnitudes in the $G$, $B_P$, and $R_P$ bands from Gaia EDR3 \citep{Gaia_2021_eDR3}, $J$, $H$, and $K_s$ bands from 2MASS \citep{Skrutskie_2006_2MASS}, and $W1$, $W2$, $W3$, and $W4$ bands from WISE \citep{2014yCat.2328....0C}.
We fit BT-Settl synthetic spectrum models \citep{Allard_2015_BTSettl} to the SED using the following parameters: effective temperature $T_{\rm eff}$, log surface gravity $\log{g}$, and $\log{(R_s/D)}$, where $R_s$ and $D$ are the radius and distance of the star, respectively. We assumed no interstellar extinction. We calculated the posterior probability distributions of these parameters using the Markov Chain Monte Carlo (MCMC) method implemented in the Python package \texttt{emcee} \citep{2013PASP..125..306F}. In each MCMC step, a synthetic spectrum was calculated by linearly interpolating the model grid for a given set of parameters, where the metallicity value was randomly chosen from a normal distribution of $\mathcal{N}(-0.20, 0.20)$~dex. 
A white noise jitter term, $\sigma_{\rm jitter}$, was also fitted for each of the Gaia EDR3, 2MASS and WISE data sets such that the magnitude uncertainty was given by $\sqrt{\sigma_{\rm cat}^2 + \sigma_{\rm jitter}^2}$, where $\sigma_{\rm cat}$ is the catalogued uncertainty in magnitude. 
From the posteriors, we derived $T_{\rm eff} = 3071^{+34}_{-22}$~K, $\log g = 5.26^{+0.18}_{-0.35}$ (cgs), and $\log(R_s/D) = -9.3721^{+0.0062}_{-0.0085}$ (cgs). Adopting $D=11.2183 \pm 0.0035$~pc from \citet{2021AJ....161..147B} which is estimated based on the GaiaEDR3 parallax, we obtained $R_s = 0.2111^{+0.0030}_{-0.0041}$~$R_\odot$, which also yielded the stellar luminosity of $L_s = 3.584^{+0.067}_{-0.071} \times 10^{-3}$~$L_\odot$ via the Stefan-Boltzmann law. Note that the median values of the white noise jitter terms are 0.089, 0.074, and 0.00050 mag for the Gaia EDR3, 2MASS, and WISE data sets, respectively. The relatively large jitter values in the Gaia EDR3 and 2MASS data sets might reflect the challenges for the stellar models for cool stars.\par

Based on the stellar metallicity reported in \citet{2022AJ....163...72I}, the effective temperature derived above, and the parameters in the literature (i.e., the Gaia parallax and 2MASS magnitudes), we inferred the physical parameters of \target, including the stellar mass, which is required to estimate the planet mass. We made use of the empirical formulae by \citet{2015ApJ...804...64M} and \citet{2019ApJ...871...63M} for the stellar radius and mass, for which the apparent $K_s-$band magnitude of $m_{K_s}=8.279\pm0.023$ mag was adopted from the 2MASS catalog. We implemented a Monte Carlo simulation to estimate the uncertainties of the output parameters, accounting for the statistical error of the input parameters as well as the systematic error of the empirical formulae. We obtained a stellar radius and mass of $0.2113 \pm 0.0063\,R_\odot$ and $0.1774 \pm 0.0045\,M_\odot$, respectively, which yield a mean stellar density of $26.5_{-2.2}^{+2.5}$ g cm$^{-3}$ and a surface gravity of $\log g = 5.038 \pm 0.027$ (cgs).  This surface gravity is consistent with that derived from BT-Settl model atmospheres.  

\subsection{Adaptive optics imaging}
\begin{figure}
    \centering
    \includegraphics[scale=1.0]{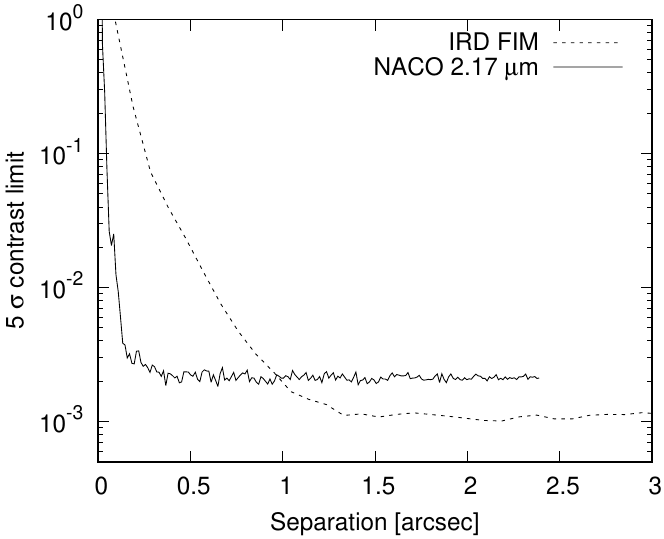}
    \caption{
    5$\sigma$ contrast limits around \target. The solid line shows {the contrast limit from VLT$/$NACO in} the NB 2.17 filter. The dashed line shows the contrast limit {from} the IRD FIM camera, which has a sensitivity {over} wavelengths ranging from 0.83 to 1.05 $\mathrm{\mu m}$
    }
    \label{fig:PSF}
\end{figure}

\target has a relatively high renormalized unit weight error (RUWE) of 1.48 in Gaia EDR3, implying that this star might be associated with an unseen companion. 
In order to search for a possible companion, we analyzed adaptive optics high-resolution images of \target\ obtained with the Fiber Injection Monitor (FIM) camera of IRD. 
FIM is an AO-assisted CCD camera sensitive to wavelengths of 0.83 to 1.05 $\mathrm{\mu m}$, and is used to monitor a target's position during observations. The CCD camera is usually used to feed the light into the IRD fiber. 
The FIM observations were performed every time just before RV measurements of IRD, but we selected images taken only under good seeing conditions. The final selected images are consist of 33 frames with a total integration time of 74 seconds. 
The FWHM of the final combined Point Spread Function (PSF) is 0\farcs19, and the 5$\sigma$ raw contrast limit is shown in Figure~\ref{fig:PSF}. 
We also processed archival VLT/NACO data for Ross 508 (program ID: 71.C-0388(A), PI: J.-L, Beuzit) obtained with a narrow-band filter at 2.17 $\mathrm{\mu m}$ (NB 2.17 filter, 2.166 $\pm$ 0.023 $\mathrm{\mu m}$) using a well-tested general-use pipeline \citep{2011ApJ...729..128C}.
A total of 44 frames with an integration time of 2 seconds each were reduced and combined to create the final high-quality image. No speckle subtraction techniques were applied to either the FIM or NACO images.
We found no stellar companions {at a separation wider than} $\sim$0\farcs{1} from the central star. 
At separations exterior to 0\farcs25 ($r_{\rm proj}$ $\sim$ 2.8 au), {the comparison of the contrast limits with the \cite{2003A&A...402..701B} evolutionary models enables us to} rule out companions that are more massive than 35 $M_{\rm J}$ or 70 $M_{\rm J}$ for an assumed system age of 1 Gyr or 10 Gyr, respectively.

\subsection{Radial Velocity and Orbital Solutions}
\subsubsection{Radial Velocity Measurements}
\label{rv_mesurements}

For each wavelength-calibrated spectrum, we measured precise RVs following the standard RV-analysis pipeline for IRD \citep{2020PASJ...72...93H}; we refer to that paper for a detailed discussion.  In short, the pipeline extracts the instantaneous instrumental profile (IP) of the spectrograph from the simultaneously injected LFC spectrum, and generates an IP-deconvolved, telluric-free template spectrum for the target star using multiple IRD spectra. Using this template, individual spectral segments for each IRD spectrum are fitted by the forward modeling technique, in which telluric absorption features are simultaneously optimized. The resulting relative RV values as well as their uncertainties are summarized in Table~\ref{tbl:rv_obs}. The RV internal error was typically 2--3 $\mps$ for each frame.

We corrected for RV drifts that are attributed to the Earth's rotation and orbital motion (i.e., barycentric RV correction) using the TEMPO2 software \citep{tempo2}. TEMPO2 also corrects for perspective acceleration, which is $\approx$0.45~m\,s$^{-1}$\,yr$^{-1}$ for \target. 
IRD applies multiple readouts to its two H2RG detector during an exposure \citep{2018SPIE10702E..60K}.
Accordingly, we computed the telluric RV using the time when half of the total signal was counted, which was determined by monitoring the photon counts acquired by the detectors every $\sim$1.5 seconds.

We note that one of the causes of long-term RV measurement instability originates from the IRD instrument. 
We evaluated the instrumental error via both laboratory experiments and on-sky monitoring observations of an RV standard star, GJ 699. These two methods resulted in the same value of 2 $\mps$. 
From the laboratory experiments, we found that the main sources of instrumental error are the intra- and inter-pixel sensitivity variations of the detector ($0.96$ $\mps$), as well as the modal noise ($\sim 1.2$ $\mps$) caused by PSF instability \citep{2018SPIE10702E..11K}.  
We found a total RV error of $3\,\mps$ over 718 days of on-sky monitoring observations of GJ 699, which (assuming {no planet around the star}) yields an instrument-derived error of about 2 $\mps$ (Kotani et al. in prep.). 
In the case of \target, we assume that the RV measurements are affected by the same amount of instrumental noise. {Note that Table \ref{tbl:rv_obs} provides RV uncertainties that do not include the instrument-derived errors}.

\subsubsection{Orbital Solutions}

\begin{figure}
    \centering
    \includegraphics[scale=0.4]{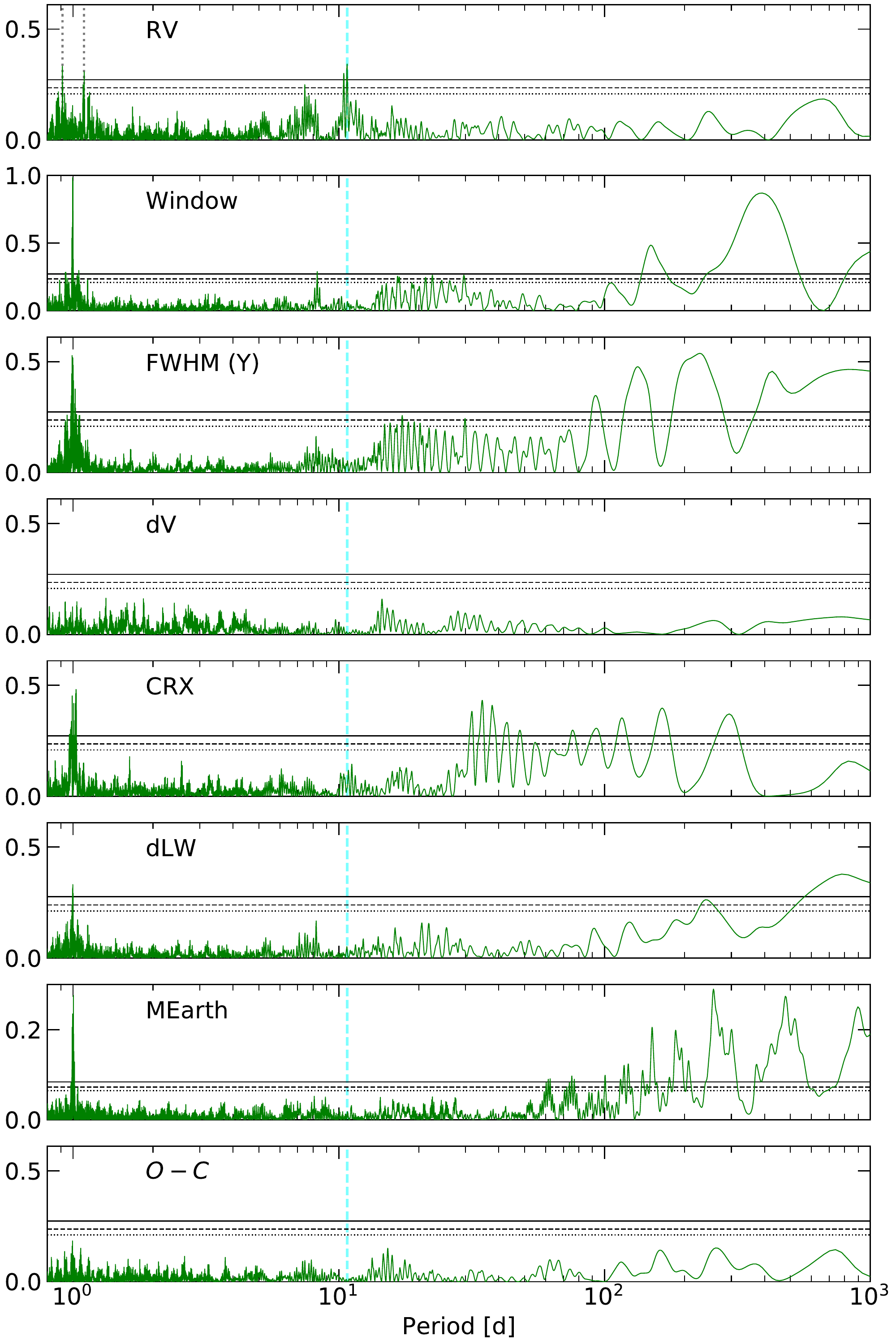}
    \caption{
    From top to bottom: GLS periodograms of the RV time series, window function of the RV sampling, FWHM in Y-band region, dV, CRX, dLW, MEarth photometry, and residual of the RV fitting. The dashed cyan vertical line indicates the RV period at 10.7510 days, while the vertical gray lines show the estimated aliases. See the main text for details of those values. The FAP values corresponding to 0.1\%, 1\%, and 5\% are shown with solid, dashed, and dotted lines, respectively.
    }
    \label{fig:gls_all}
\end{figure}

We searched for periodicity in our RV time series before performing an orbital fit.  
We computed the Generalized Lomb-Scargle (GLS) periodogram \citep{2009A&A...496..577Z} for all RV data and for the window function, and identified three significant peaks at $10.7510$, $0.9124$, and $1.1023$ days in order of decreasing GLS power (Figure~\ref{fig:gls_all}).
Hereafter, all the FAP values were derived by analytical estimation reported in \citet{2008MNRAS.385.1279B}.
For the window function, we identified a single peak at $0.9972$ days.
{In the GLS periodogram analysis, we set the RV error to be $\sqrt{\sigma_{i}^2 + \sigma_{\rm{inst}}^2}$, where $\sigma_{i}$ is the RV uncertainty of an $i$-th observation and $\sigma_{\rm{inst}}$ is the instrument-derived error described in Section \ref{rv_mesurements}}.

Although the RV periodogram shows several significant periodicities, we first investigate whether some of these represent cases of aliasing, which generally appears in periodograms of discretely sampled time series data.
To distinguish aliases from physical signals, we performed a simple alias analysis based on the computed periodograms.
In general, when sampling a sine wave of frequency $f$ at sampling frequency $f_{s}$, the sample is indistinguishable from any other sample of the sine curve whose frequency is $f_{\mathrm{signal}}\left(N\right) = |f-Nf_{s}|$ (where $N=0,\pm1,\pm2,\cdots$, and $f_{\mathrm{signal}} \left(0\right) = f$ is the actual signal frequency) as they yield identical sets of data. 
Those frequencies other than $N=0$ are aliases that should be addressed.
This equation assumes $f_{s}$ as a perfectly evenly spaced sampling, and of course, the actual observations will not be carried out with such an ideal interval.
However, since the window function of our data sampling shows a dominant power on almost a single frequency, we should be able to estimate the approximate effect of aliasing by applying this equation.
We here assumed the most significant RV frequency of $1/10.7510~\mathrm{d^{-1}}$ to be a physical one and the sampling frequency to be the most significant window function peak of $1/0.9972~\mathrm{d^{-1}}$.
If $N=+1$ and $N=-1$, this yields $1/1.0992~\mathrm{d^{-1}}$ and $1/0.9125~\mathrm{d^{-1}}$ respectively. 
These two frequencies are almost identical to those of the second and third significant peaks of the periodogram, showing that those two peaks in the periodogram can be interpreted as alias phenomena associated with a period of $10.751$ days and its dominant sampling interval of $0.9972$ days.
We note that if we assumed the secondary peak of the window function at $390.25$ days as a sampling frequency, the aliases were $11.05$ and $10.46$ days {in the case of} $N=+1$ and $N=-1$, respectively.
The $10.46$ days alias is almost identical to a peak of $10.45$ days in the periodogram though its frequency is far beyond the Nyquist frequency of $0.5f_{s}=1/780.5$.

With a single significant periodicity at 10.75~days, we next performed a Keplerian fit to the RVs. 
As discussed in Section \ref{sec:activity_analysis}, we found no significant activity signals at this period.  
{We used \texttt{emcee} to explore the parameter spaces via MCMC}.
The initial states were randomly generated from the prior distributions shown in Table \ref{tab:oc_results}.
We ran the sampler until it satisfied the following convergence criterion: if the number of steps is greater than $100$ times the autocorrelation length of each parameter, which is estimated every $1000$ steps, and this estimate varies by less than $1\%$, then we assume that the chain has converged. The maximum steps was set to 30 millions.
The first $20\%$ of the steps were discarded as burn-in, yielding a total of $24.0$M samples of the posterior distribution from the remaining steps.

Based on \citet{2005ApJ...631.1198G}, the likelihood function $\mathcal{L}$ used in this analysis is
\begin{equation}
        \ln \mathcal{L} = -\frac{1}{2}\sum_{i}\left(\frac{\left(v_{i,\mathrm{obs}}-v_{i,\mathrm{model}}\right)^2}{\sigma_{i}^2+\sigma_{\mathrm{jitt}}^2}
    + \ln {\left(\sigma_{i}^2+\sigma_{\mathrm{jitt}}^2\right)}\right),
\end{equation}
where $v_{i,\mathrm{obs}}$ is the $i$-th observed RV, $v_{i,\mathrm{model}}$ is $i$-th RV model calculated from the companion's Keplerian orbital motion, and $\sigma_{\mathrm{jitt}}$ is a jitter parameter to account for RV variations due to stellar activity and changes in instrumental stability.
Priors of the parameters, best-fit orbital solutions and their uncertainties are presented in Table~\ref{tab:oc_results}, where $K_b$ is the velocity semi-amplitude, $P$ is the orbital period, $T_p$ is the time of periastron passage, $e$ is the eccentricity, $\omega$ is the argument of periastron, $\gamma$ is the constant velocity, and $\dot{\gamma}$ is the constant RV acceleration (i.e. ~linear RV trend).
The $M_b,~i$ and $a_b$ denote the mass of the planet, orbital inclination relative to line-of-sight and its orbital semimajor axis, respectively.\par
A relatively large offset in RV measurements appears on August and September of 2021 (See Fig. \ref{fig:rv_fit_A1}).
We wondered if these observations were influenced by a possible irregular and temporal offset in our RV measurements possibly caused by an instability of the instrument or a high activity event such as flaring.
Indeed, although the LFC's spectra have been stabilized for several years, the observing runs at August and September 2021 were immediately after the restoration from the irregular operation of the temperature stabilization room in which the LFC instrument are placed. 
Therefore, we compared two RV models: A) one does not consider the RV offset in this period, B) another assumes the RV offset as an additional systemic RV offset parameter in the RV model ($\gamma_{2}$).

\begin{table*}[]
    \centering
    \tbl{RV posterior distributions and priors. \footnotemark[$*$]}{
    \begin{tabular}{lllllcc}
        \hline \hline
        Parameter & Model A1 & Model A2 & Model B1 & Model B2 & Prior & Bound \\
        \hline
        $K_b~(\mps)$ & ${3.80}^{+0.62}_{-0.58}$ & ${3.90}^{+0.67}_{-0.61}$ & ${3.92}^{+0.60}_{-0.58}$ & ${3.95}^{+0.63}_{-0.59}$ & Uniform & $\left(0, 10\right)$ \\
        $P$ (d) & ${10.76}^{+0.01}_{-0.02}$ & ${10.77}^{+0.01}_{-0.01}$ & ${10.77}^{+0.01}_{-0.01}$ & ${10.77}^{+0.01}_{-0.01}$ & Uniform & $(9, 12)$ \\
        $T_p$ (BJD $-2450000$) & ${9370.22}^{+1.38}_{-1.88}$ & ${9370.63}^{+0.80}_{-0.84}$ & ${9370.24}^{+0.64}_{-0.71}$ & ${9370.34}^{+0.65}_{-0.60}$ & Uniform & $2459370 + (-6, +6)$ \\
        $e$ & $0$ $(<0.70;~3\sigma$) & ${0.33}^{+0.15}_{-0.17}$ & ${0.33}^{+0.13}_{-0.15}$ & ${0.36}^{+0.14}_{-0.16}$ & Uniform & $(0, 0.99)$\\
        $\omega$ (rad) & & ${-0.41}^{+0.49}_{-0.50}$ & ${-0.65}^{+0.39}_{-0.48}$ & ${-0.63}^{+0.38}_{-0.44}$ & Uniform & $(-\pi, +\pi)$\\
        $\sigma_{\mathrm{jitt}}~(\mps)$ & ${2.52}^{+0.43}_{-0.41}$ & ${2.26}^{+0.44}_{-0.42}$ & ${1.76}^{+0.46}_{-0.48}$ & ${1.73}^{+0.48}_{-0.49}$ & Uniform & $(0,20)$ \\
        $\gamma~(\mps)$ & ${-0.65}^{+0.38}_{-0.38}$ & ${-0.28}^{+0.38}_{-0.38}$ & ${-0.19}^{+0.34}_{-0.34}$ & ${-0.08}^{+0.35}_{-0.35}$ & Uniform & $(-20,+20)$ \\
        $\gamma_{2}~(\mps)$ &&& ${-7.68}^{+1.26}_{-1.31}$ & ${-6.61}^{+1.44}_{-1.36}$ & Uniform & $(-20,+20)$ \\
        $\dot{\gamma}~(\mps{\mathrm{yr}^{-1}})$ && ${2.03}^{+0.62}_{-0.62}$ & & ${0.89}^{+0.62}_{-0.60}$ & Uniform & $(-10,+10)$ \\
        $M_b\sin i~(\mear)$ & ${3.99}^{+0.60}_{-0.60}$ & ${3.96}^{+0.59}_{-0.59}$ & ${4.00}^{+0.53}_{-0.55}$ & ${3.97}^{+0.54}_{-0.58}$ &  &  \\
        $a_b$ (au) & ${0.05353}^{+0.00047}_{-0.00051}$ & ${0.05361}^{+0.00047}_{-0.00048}$ & ${0.05366}^{+0.00056}_{-0.00049}$ & ${0.05356}^{+0.00048}_{-0.00056}$ &  &  \\
        $\mathrm{rms}~(\mps)$ & & ${3.73}$ & ${3.29}$ & ${3.27}$ & & \\
        \# of Samples & $16.0$\,M & $13.8$\,M & $24.0$\,M & $20.5$\,M & & \\
        AIC & $582.63$\footnotemark[$**$] & $573.79$ & $553.63$ & $553.44$ &&\\
        BIC & $600.94$\footnotemark[$**$] & $602.56$ & $582.39$ & $584.82$ &&\\
        Description & Single planet & Model A1 & Single planet & Model B1 &&\\
        &&$+$ Linear Trend & $+\gamma_{2}$ &  $+$ Linear Trend &&\\
        \hline
    \end{tabular}}
    \label{tab:oc_results}
    \begin{tabnote}
    \hangindent6pt\noindent
    \hbox to6pt{\footnotemark[$*$]\hss}\unskip%
    The $e$ and $\omega$ were derived from $\sqrt{e}\sin\left(\omega\right)$ and $\sqrt{e}\cos\left(\omega\right)$. \\
    \hbox to6pt{\footnotemark[$**$]\hss}\unskip%
    Assumed $e=0$ as the best-fit value.
    \end{tabnote}
\end{table*}
We report the posterior median and $1\sigma$ credible region for each parameter in Table~\ref{tab:oc_results}.
In our analysis, we compared four RV models in total:

\begin{itemize}
    \item A1) Single planet,
    \item A2) Model A1 $+$ linear RV trend,
    \item B1) Single planet with RV offset for data in August and September of 2021 (i.e. $\gamma_{2}$), and 
    \item B2) Model B1 $+$ linear RV trend.
\end{itemize}
To perform model selection, we calculated the Akaike Information Criterion \citep[AIC;][]{akaike_1974} and Bayesian Information Criterion \citep[BIC;][]{bic_1978} for each of the four models, which are defined as
\begin{equation}
    \mathrm{AIC} = 2k-2\ln \mathcal{L}
\end{equation}
and
\begin{equation}
    \mathrm{BIC} = k\ln\left(N\right)-2\ln \mathcal{L},
\end{equation}
respectively, where $k$ is the number of parameters, $N$ is the number of data points, and $\mathcal{L}$ is the maximum likelihood of the model.

\begin{figure}
    \centering
    \includegraphics[scale=0.47]{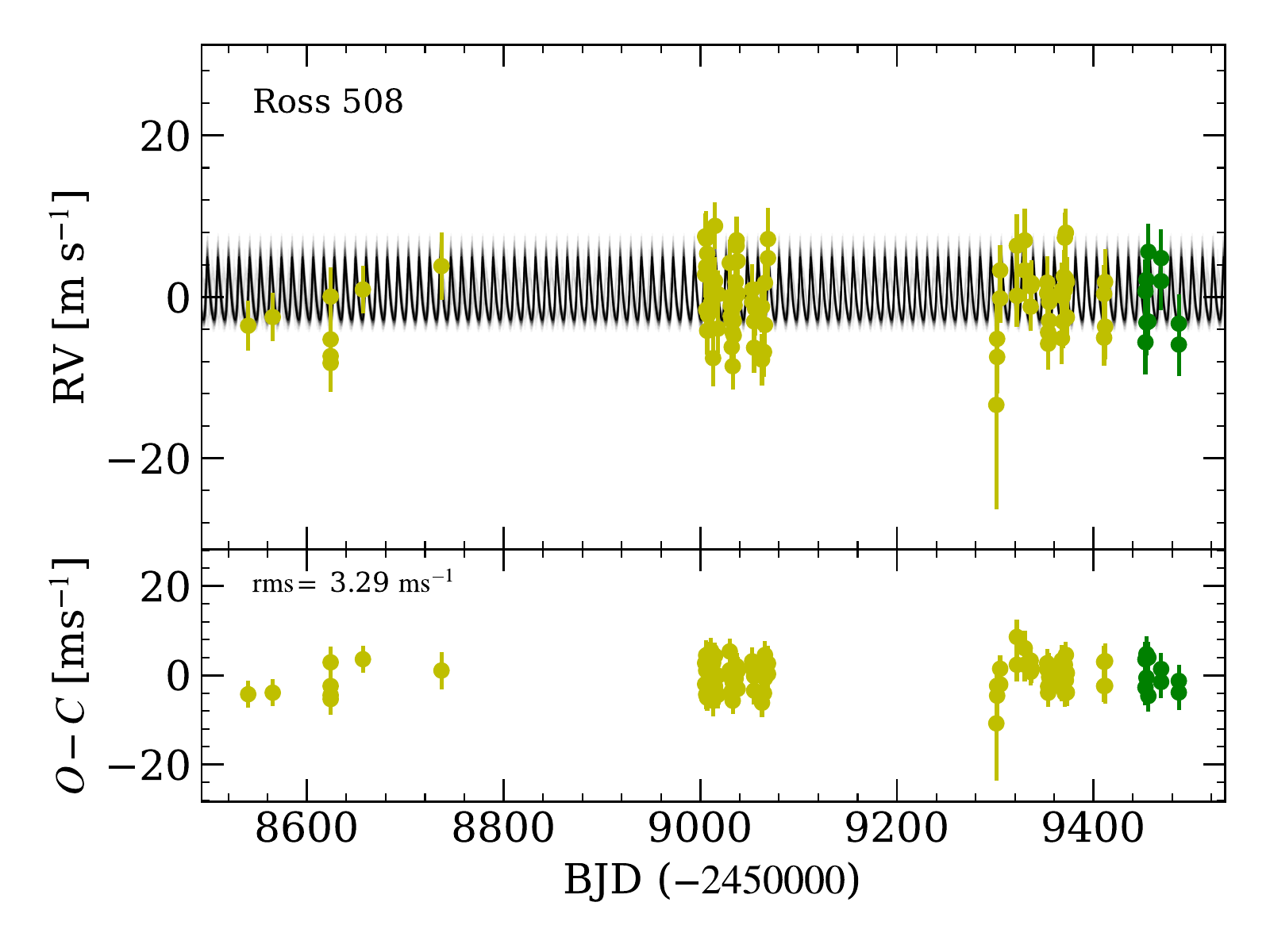}\\
    ~~\includegraphics[scale=0.45]{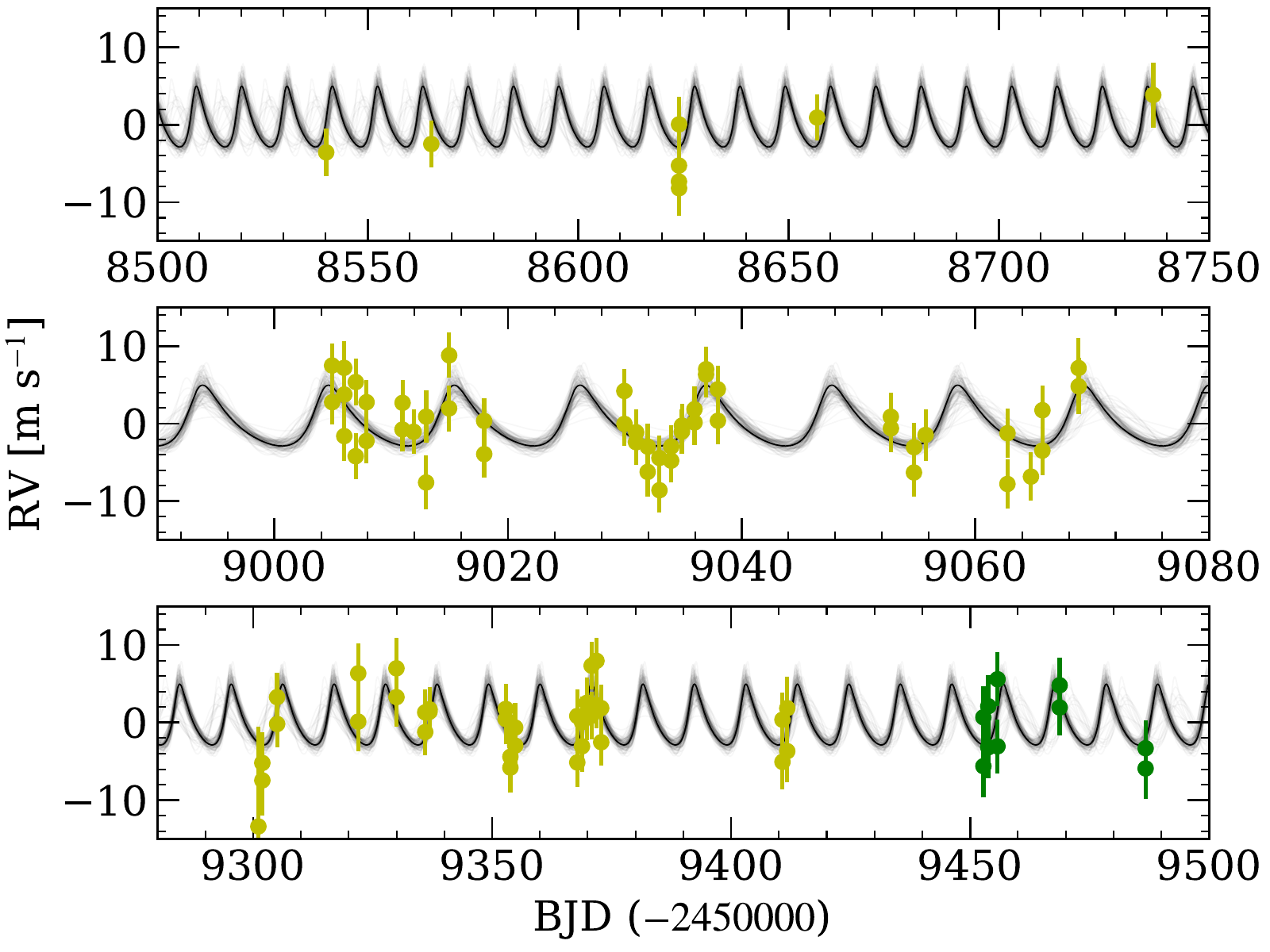}\\
    \includegraphics[scale=0.45]{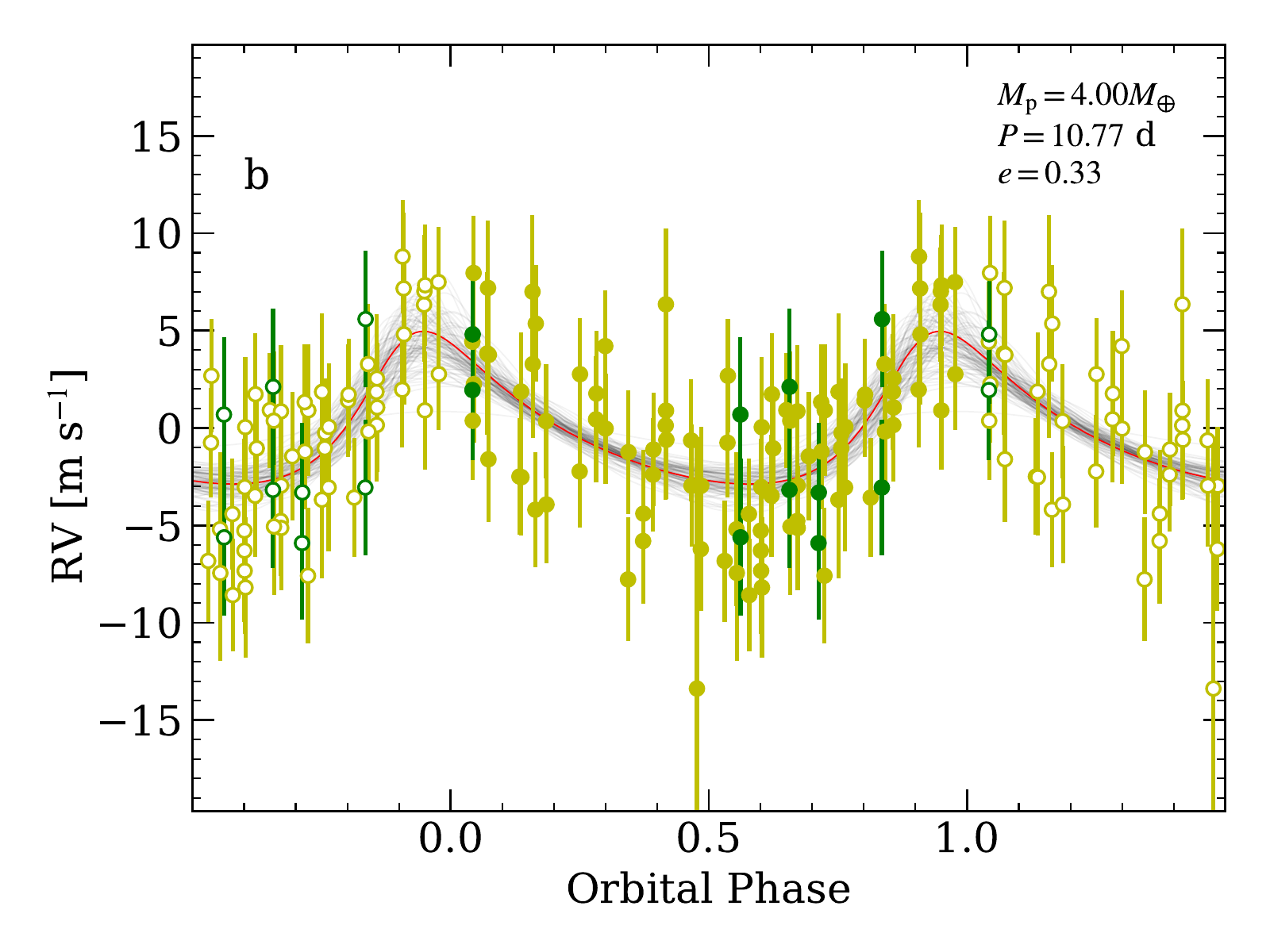}
    \caption{Observed RVs and the best-fit orbital solution based on the MCMC method of Model B1 (red line). Yellow circles with error bars represent the observed RVs and {their} uncertainties, respectively. Green circles and error bars show the same as yellow ones, but obtained in August and September 2021. Gray lines represent 100 RV model curves randomly selected from the posterior. The error bars are the quadrature sum of the measurement uncertainties with the best-fit value of the jitter parameter $\sigma_{\mathrm{jitter}}$. \textit{Top}: the best-fit RV curve (top) and the residuals of the fit (bottom). \textit{Middle}: Zoom up of each data segment. \textit{Bottom}: Phase-folded RV variations. Red line represents the best-fit RV model.}
    \label{fig:rv_fit_B1}
\end{figure}

\begin{figure*}
    \centering
    \includegraphics[scale=0.25]{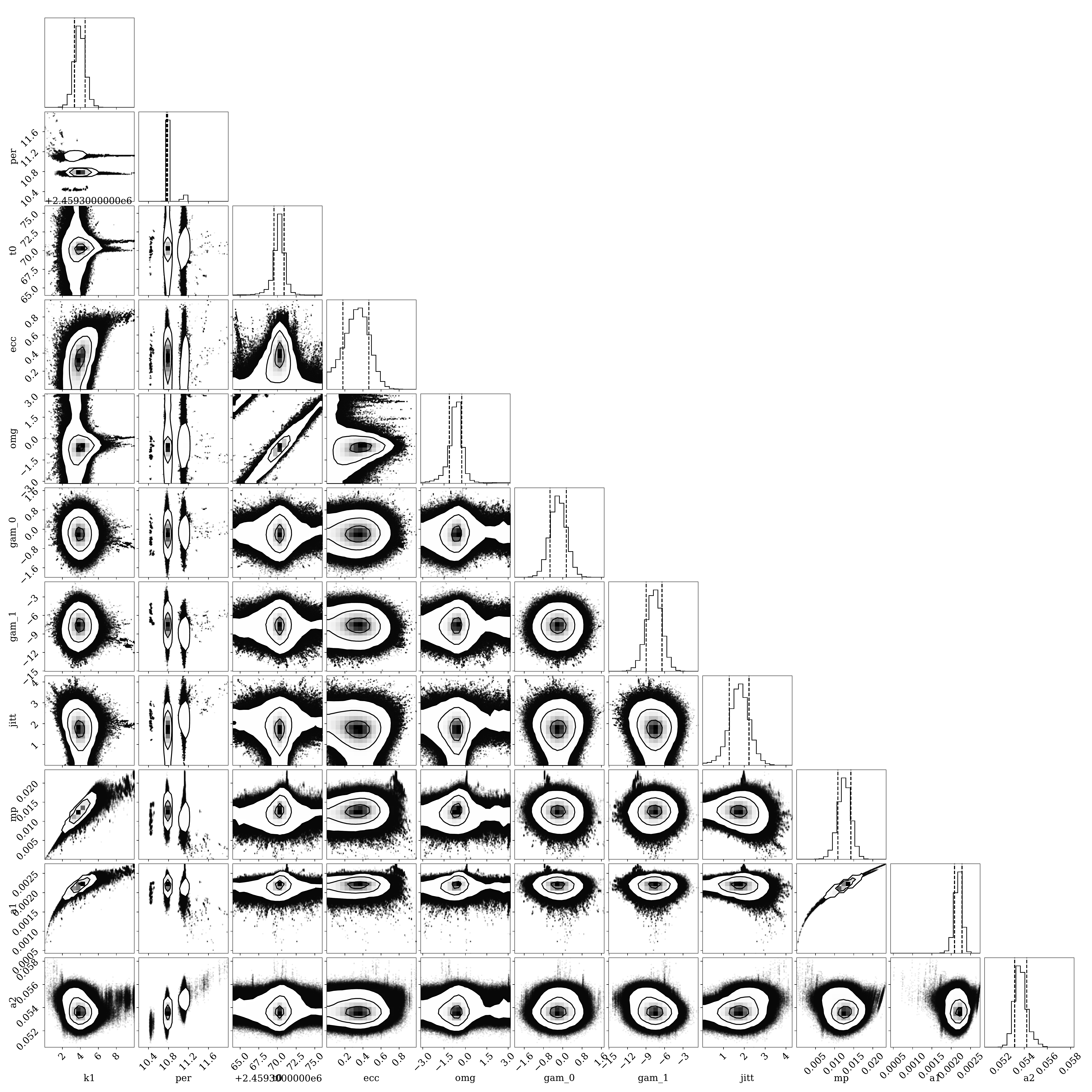}
    \caption{Corner plot of the posterior distributions of Model B1.}
    \label{fig:corner_B1}
\end{figure*}

Figure~\ref{fig:rv_fit_B1} shows the observed RVs and the orbital solutions from our MCMC analysis, and Figure~\ref{fig:corner_B1} shows a ``corner'' plot of the covariance between the parameters in our MCMC analysis.
The BIC value is smaller for the model A1 than that of the model A2 while the AIC value is not.
The $\Delta AIC$ and $\Delta BIC$ are only slightly different from the values at which a model selection is statistically meaningful \citep{kass1995bayes}.
We also found no clear statistical difference in the comparison between the B1 and B2, as indicated by the comparable BIC values.
While Model A2 suggests a long-term trend, Model B2 did not, suggesting a degeneracy between the models of the linear RV trend and the temporal RV offset.
Further investigation and additional data 
can resolve the degeneracy, but we here conclude that there is no clear evidence to identify a long-term linear trend in our RV measurements.\par
The posterior eccentricity distribution of the model A1 monotonically decreases with a maximum at zero (Figure~\ref{fig:corner_A1} in the Appendix); we report only the $3\sigma$ upper limit for the eccentricity.
Meanwhile, the posterior distributions of eccentricity for models A2, B1, and B2 have a maximum likelihood value around $0.3$. 
However, a zero eccentricity is still likely as indicated by the eccentricity posterior in these three models.
We therefore conclude that only an extremely high-eccentricity, $e > 0.9$ ($3\sigma$), is unlikely.
Furthermore, we adopt the model B1 (i.e., inclusion of no linear RV trend and a systematic RV offset) as our fiducial model based on the lowest AIC/BIC value among the four models. 

\subsection{Stellar Activity}
\label{sec:activity_analysis}
\subsubsection{Photometric Variability}

While the RV data are well-fit by a planetary companion, we now assess whether stellar activity could instead be responsible. 
To search for photometric modulation caused by stellar surface magnetic activity, we used the public archive data from the MEarth-North project \citep{2012AJ....144..145B} 
 and the ``All-Sky Automated Survey for Supernovae'' \citep[ASAS-SN; ][]{2014ApJ...788...48S}. 
 We analyzed the MEarth data for \target from Data Release 10, and we selected data observed with the same telescope and RG715 filter bandpass at MEarth-North. 
 We analyzed the g-band data from ASAS-SN.
 TESS will observe this target in Sector 51 (April to May 2022), which will allow us to characterize the stellar activity in more detail.

\citet{2016ApJ...821...93N} were not able to detect the rotation period of \target.
We independently analyzed the photometric data both from the MEarth and the ASAS-SN, and found no clear signals in the GLS periodogram that could be due to rotation or that match the observed RV signal (Figure~\ref{fig:gls_all} for the periodogram on the MEarth light curves).
In order to evaluate how small of a stellar-rotation modulation MEarth data can detect, we estimated the sensitivity of the MEarth data set to photometric modulation by creating 100 sinusoidal curves with periods of 10.8 days (same as the detected planet's orbital period), with the same cadence as the actual MEarth data.
We added white noise to each data point by sampling a Gaussian distribution with standard deviation equal to the individual photometric uncertainty of the corresponding data point, after scaling the median of the uncertainties to the standard deviation of all the data points in the MEarth light curve.
We then applied a periodogram analysis to these mock data.
We repeated the above analysis varying the amplitudes of the sinusoidal curves.
As per our definition, a periodic signal at 10.8 days can be detected if the False Alarm Probability (FAP) at that period is less than 1\%.
We found that 70\% of the simulations yield a detection of the 10.8-day sinusoidal signal if its amplitude is larger than 0.4\,\% of the stellar brightness in the MEarth photometric band. 
Similar results were obtained even if we shifted the phase of the sinusoidal signal or directly injected the sinusoidal curves into the MEarth light curves instead of creating mock data.
However, even if there were a cool spot that produces a light-curve variation equal to or smaller than 0.4\,\%, the corresponding RV semi-amplitude would be too small to account for our detected RV amplitude. 
Assuming the star's effective temperature, spot temperature, and rotation velocity ($v\sin{i}$) to be 3000 K, 2500 K, and 1 km\,s$^{-1}$, respectively, such a cool spot would induce an RV semi-amplitude of no more than 2 m\,s$^{-1}$.
Here, the $v\sin{i}$ of 1 km\,s$^{-1}$ is the maximum value estimated from the stellar radius of 0.213 $R_{\odot}$ (see Table \ref{tab:stellarparam}),
assuming a rotation period equal to the detected RV period.
Thus, a cool spot rotating with a period of 10.8 days and covering an area smaller than 0.4\% of the stellar surface cannot reproduce our identified RV variation.      
Our light-curve analysis suggests that the 10.75 days signal is not caused by a cool spot on the stellar surface.
We note that this analysis only applies if the phase of the photometric modulation is coherent over the $\approx$3 year baseline of the IRD observations.  However, the same criterion applies to the RV signal itself, which is indeed coherent in phase over this baseline.
 
\subsubsection{Line Profile Variation}

We next determine whether there is periodicity in the line profile at a period matching that of our recovered planet. 
We apply the least-squares deconvolution (LSD) method \citep{2010AA...524A...5K} to derive mean line profiles.
A list of lines is empirically built from an IP-deconvolved, telluric-free template spectrum.
To minimize contamination, we use spectra within 1000--1070 nm, which contain fewer telluric lines.
The uncertainties of the LSD profiles are determined with formal uncertainties scaled by the standard deviations of the difference between each individual LSD and the mean profile.
As indicators of line profile variation, we computed the full width at half maximum (FWHM; the line width) and BiGauss (dV; the line asymmetry) by fitting Gaussian functions \citep{2015MNRAS.451.2337S}.
We also computed the chromatic index (CRX; the wavelength dependence of RV) and the differential line width (dLW; the line width) \citep{2018AA...609A..12Z}.
To compute the dLW, we used a template spectrum convolved with the averaged IP instead of a coadded spectrum to avoid telluric-line contamination.
To determine the CRX, the wavelength range is binned from $1000~\mathrm{nm}$ to $1750~\mathrm{nm}$ in $10~\mathrm{nm}$ increments, and we use the weighted average of the RVs of the segments in each bin.

\begin{figure}
    \centering
    \includegraphics[scale=0.5]{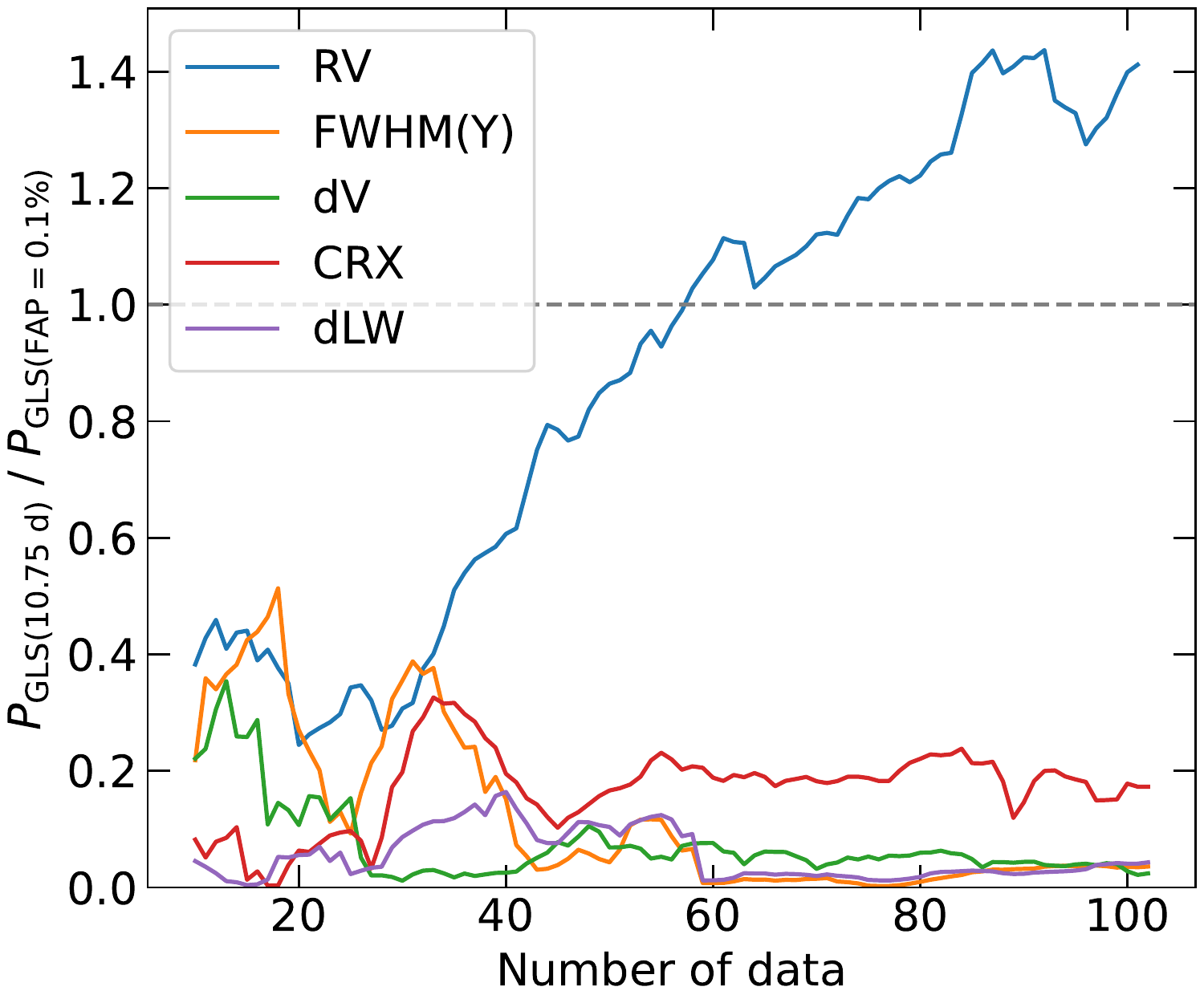}
    \caption{Significance of the periodogram power for RVs and for all line-profile indexes as the number of data points {increases}.  Significances are measured as fractions of the GLS power corresponding to FAP$=0.1\%$; the dashed gray horizontal line shows a value corresponding to FAP$=0.1\%$ (i.e. $1.0$).
    The powers for RVs, FWHM(Y), dV, CRX, and dLW are respectively shown in blue, orange, green, red, and purple.
    }
    \label{fig:period_evo}
\end{figure}

The GLS periodograms of all stellar activity indicators are shown in Figure~\ref{fig:gls_all}.
None of the activity indicators exhibit any significant peaks at 10.75 days.
Figure~\ref{fig:period_evo} shows the evolution of GLS power at 10.75 d for the RVs and activity indicators.
While the power increases with the number of data points for the RVs,
the power remains consistently low for the activity indicators, {suggesting that} the periodic RV variations are not induced by stellar activity \citep[e.g.][]{2017AA...601A.110M}.

\section{Summary and Discussion}\label{Discussion}

In the previous section, we showed that the M4.5 dwarf \target\ has a significant RV periodicity at 10.75 days with possible aliases at 1.099 and 0.913 days. 
This periodicity has no counterpart in photometry or stellar activity indicators, but is well-fit by a Keplerian orbit due to a new planet, \target\ b. 
Our newly discovered planet, \target\ b, has a minimum mass of 4.0 $\mear$ and {a semi-major axis of} 0.05 au.

We explored four possible scenarios to explain the measured RV data. We examined models including a presence of RV offset to the data obtained in August and September 2021, and a long-term RV trend, which might be caused by an unseen companion, because \target\ has a relatively high renormalized unit weight error (RUWE) of 1.48 in Gaia EDR3, suggesting that it is poorly fit by a single star model. Of 19 comparison stars in EDR3 with parallaxes between 80 and 100~mas and $B_P - R_P$ colors within 0.3 mag of that of \target, just three have RUWE values higher than 1.48.

While the differences between the four models are not large, we found that a $\sim$ 7 $\mps$ RV offset and the absence of a long-term RV trend best explain the observed data. In this scenario, the peak of the posterior distribution of the eccentricity is around 0.3, but the distribution is wide all the way down to zero; hence it does not constrain the eccentricity well. As a reference, some previously known exoplanets around late-M dwarfs have eccentricities reported as upper limits, such as GJ 1061 b ($e < 0.31$), GJ 1061 d \citep[$e<0.53$;][]{2020MNRAS.493..536D} and Proxima Centauri b \citep[$e<0.35$;][]{2016Natur.536..437A,2017ApJ...844..100B}.
Further RV measurements of \target will clarify whether the planet {has a} high eccentricity among the sample of known super-Earths around mid-to-late M stars $(T_{\mathrm{eff}} < 3200~\mathrm{K},~M_p\sin i < 10\mear)$, providing important clues about their origin. 

As well as other super Earths with orbital periods much shorter than {the snow line around their host stars}, \target\,b may have formed beyond the snow line {($\sim 0.16$\,au)} and undergone inward Type I migration \citep{1979ApJ...233..857G,2009ApJ...699..824O,2017MNRAS.470.1750I}.
Even if the eccentricity of a migrating planet is initially high, it can be damped by the force exerted on the planet by density waves \citep[e.g.,][]{2004ApJ...602..388T}.
Thus, the solution of a single-planet system with zero or low eccentricity is compatible with the Type I migration scenario.
Alternatively, there remains the possibility that \target\,b is in a high eccentricity orbit.
In a multiple-planet system, migrated planets experience giant impacts or are trapped in a resonant chain \citep[e.g.,][]{2009ApJ...699..824O,2017MNRAS.470.1750I}. Planetary eccentricities are excited by giant impacts.
The eccentricity of a planet can be also excited by gravitational interactions between neighboring planets or secular perturbations from a (sub)stellar companion {on} a wider orbit.
The confirmation of a long-term RV trend will help disentangle the formation history of the super-Earth \target\,b.

\begin{figure}[h]
 \begin{center}
  \includegraphics[width=80mm]{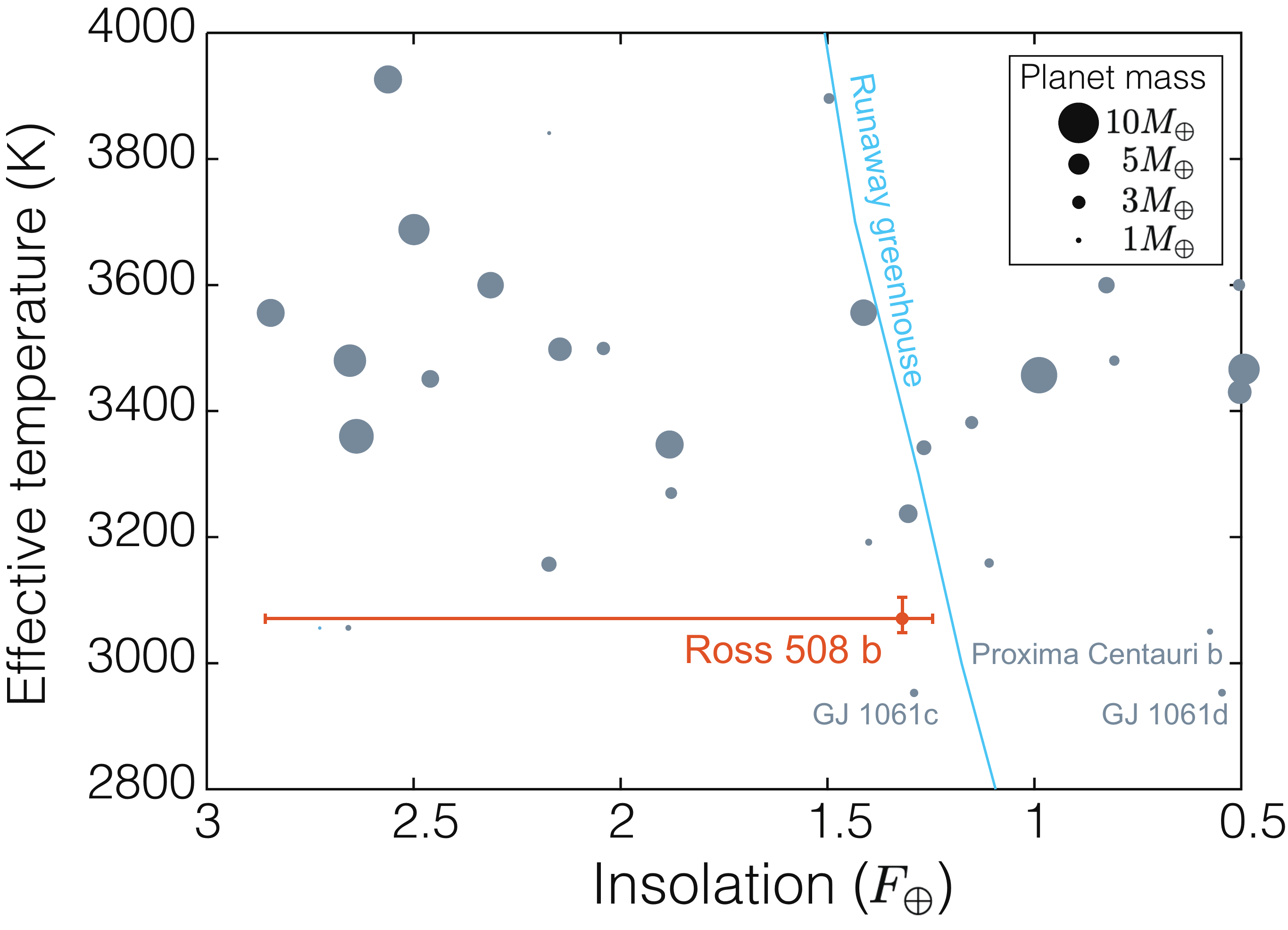}
 \end{center}
 \caption{Planets with masses below 10 $M_\oplus$ around stars with $2800$\,K $\leq T_\mathrm{eff} \leq 4000$\,K. \target\,b is located near the runaway greenhouse limit as the inner edge of a habitable zone for an Earth-sized aquaplanet around M dwarfs \citep{2017ApJ...845....5K}.
 The error bar of Ross 508's insolation is based on the $3\sigma$ limit on its eccentricity. 
 }
 \label{fig:HZ}
\end{figure}

The habitability of a planet primarily depends on the time-averaged stellar flux $\langle F \rangle$ that it receives over an entire orbit \citep{2002IJAsB...1...61W}: $\langle F \rangle = F/\sqrt{1-e^2}$, where $F$ is the stellar flux at the semimajor axis of a planet and $e$ is the eccentricity of a planet. 
As shown in Figure~\ref{fig:HZ}, the average insolation of \target\,b with an eccentricity ranging from 0 to 0.9 (which corresponds to 3 $\sigma$ limit) is always higher than the runaway greenhouse limit for an Earth-sized aquaplanet around M dwarfs \citep{2017ApJ...845....5K}. 
We note that the runaway greenhouse limit shown in Figure ~\ref{fig:HZ} was estimated for an Earth-sized planet around a low-mass star with [Fe/H]=0. 
The inner edge of the habitable zone may be farther from \target\ than what we calculated above because the low metallicity ([Fe/H] = $-$0.2) of \target\ yields a lower stellar luminosity \citep[][]{2016ApJ...819...84K}.
Also, the habitability of super-Earths can be affected by climate and mantle dynamics, such as plate tectonics \citep[e.g.,][]{2018EP&S...70..200M}.
The detailed characterization of \target helps understand the habitability of a super-Earth.

For compositional and atmospheric characterizations, it is advantageous if \target b transits the host star.
The geometric transit probability \citep[e.g.,][]{Kane2009} of \target b based on the best-fit orbital parameters (Model B1) is estimated to be $\approx 1.6\,\%$, a small probability but it is worth searching for their signals given the brightness of \target\ especially in the near infrared. We visually inspected \target's light curves by MEarth (Section \ref{sec:activity_analysis}), and found no evidence for planetary transits of \target b. Fortunately, TESS is scheduled to observe \target in Sector 51 (April to May 2022), which would deliver \target's light curve with a better precision. 
Provided that \target b has an internal composition similar to Earth, the expected depth of the transit is $\approx 0.3\,\%$, which is easily identified by the TESS photometry.
Future atmospheric characterization of \target\,b makes it possible to explore the bulk composition of \target\,b and the formation mechanism of a massive terrestrial planet orbiting near the habitable zone.

\begin{figure}[h]
 \begin{center}
  \includegraphics[width=90mm]{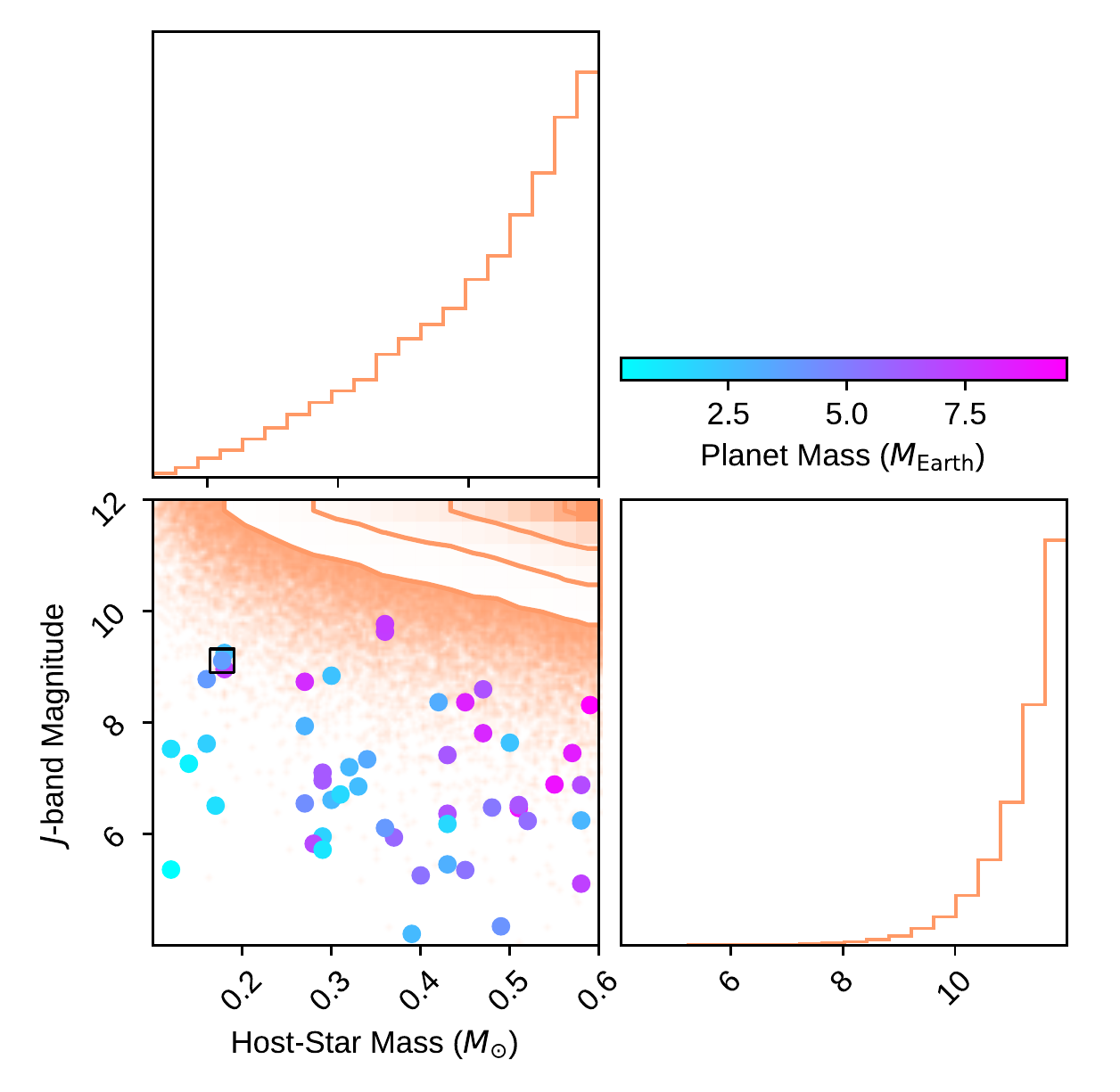}
 \end{center}
 \caption{Masses ($M_\odot$) and $J$-band magnitudes (mag) of $M$-type dwarfs that host RV-discovered planetary systems. The background contour shows the nearby $M$-type dwarfs taken from the TIC catalogue \citep{Stassun_2019}. 
 The filled circles are from the NASA exoplanet archive, with their $J$-band magnitudes from the 2MASS catalogue \citep{Skrutskie_2006_2MASS}.  
 Note that we here plot only planetary-systems hosting planets with masses lower than 10 $M_{\rm{Earth}}$. 
 \target, with parameters taken from Table \ref{tab:stellarparam}, is enclosed by a square box. The masses of the lowest-mass planets (which are also from the NASA Exoplanet Archive) in the systems are indicated by the color bar at the right side of the upper panel. A low-mass planet recently discovered around Proxima Centauri \citep{Faria_2022_Proxima} was included in the plots.
 }
 \label{fig:Mdwarf_Mass_Jband}
\end{figure}

Figure~\ref{fig:Mdwarf_Mass_Jband} places \target\ b in context with planetary systems around other nearby M-dwarfs{;} \target\ is one of the faintest, lowest-mass stars with an RV-detected planet.
{RV monitoring of} such a faint, red star requires both a large telescope aperture and a high-precision spectrograph in the near-infrared. Future surveys with IRD and other high-precision NIR spectrographs will enable the discovery of planets around {more stars like Ross 508}, and will establish the diversity of their planetary systems.
Exoplanet exploration will be advanced by the other late-M dwarf RV surveys using high-dispersion spectrographs, such as HPF, CARMENES, and SPIROU, as well as exoplanet surveys using the transit technique from space (e.g., \textit{TESS}) and the ground \citep[e.g., SPECULOOS;][]{Delrez_2018_SPECULOOS}.
Hence, the findings from various late-M dwarf observing campaigns in the 2020s will be combined to provide important clues to reveal the true nature of planetary systems around cool M dwarfs.

\begin{ack}
This research is based on data collected at the Subaru Telescope, which is operated by the National Astronomical Observatory of Japan. 
We are honored and grateful for the opportunity of observing the Universe from Maunakea, which has the cultural, historical, and natural significance in Hawaii.
M.T. is supported by JSPS KAKENHI grant Nos. 18H05442, 15H02063, and 22000005.
YH was partly supported by a Grant-in-Aid for Scientific Research on Innovative Areas (JSPS KAKENHI Grant Number 18H05439).
This work is partly supported by JSPS KAKENHI Grant Numbers JP18H05439 and JP21K20388, JST CREST Grant Number JPMJCR1761, the Astrobiology Center of National Institutes of Natural Sciences (NINS) (Grant Number AB031010).
This research has made use of the SIMBAD database, operated at CDS, Strasbourg, France \citep{2000A&AS..143....9W}.
This research made use of Astropy,\footnote{http://www.astropy.org} a community-developed core Python package for Astronomy \citep{astropy:2013, astropy:2018}.
This research has made use of the NASA Exoplanet Archive (https:\/\/exoplanetarchive.ipac.caltech.edu), which is operated by the California Institute of Technology, under contract with the National Aeronautics and Space Administration under the Exoplanet Exploration Program.
This research has made use of data obtained from or tools provided by the portal exoplanet.eu of The Extrasolar Planets Encyclopaedia.
The \texttt{corner} \citep{corner} Python module has been helpful to create the figures in this paper.

\end{ack}

\bibliography{reference}{}
\bibliographystyle{aasjournal_mod}

\clearpage
\appendix

\onecolumn
\section{Radial Velocity Measurements}
\begin{longtable}{*{3}{r}}
\caption{RVs for \target.}\\
\hline
BJD & RV & Uncertainty  \\
$(-2450000)$ & $(\mps)$ & $(\mps)$  \\
\hline
\endfirsthead
\multicolumn{3}{r}{(Continued)} \\
\hline
BJD & RV & Uncertainty  \\
$(-2450000)$ & $(\mps)$ & $(\mps)$  \\
\hline 
\endhead
\hline
\endfoot
\endlastfoot
$8540.101484$ & $-3.40$ & $2.50$ \\
$8565.084859$ & $-2.30$ & $2.43$ \\
$8623.965979$ & $-5.06$ & $4.36$ \\
$8623.971075$ & $-7.11$ & $2.71$ \\
$8623.984402$ & $0.27$ & $3.14$ \\
$8623.988127$ & $-7.98$ & $3.13$ \\
$8656.804690$ & $1.10$ & $2.37$ \\
$8736.724410$ & $4.01$ & $3.80$ \\
$9004.939392$ & $7.70$ & $2.23$ \\
$9004.946975$ & $2.95$ & $2.24$ \\
$9005.972814$ & $7.39$ & $2.99$ \\
$9005.980543$ & $-1.40$ & $2.69$ \\
$9005.988855$ & $3.97$ & $2.84$ \\
$9006.960442$ & $-3.99$ & $2.35$ \\
$9006.967935$ & $5.54$ & $2.42$ \\
$9007.882653$ & $-2.05$ & $2.30$ \\
$9007.890162$ & $2.98$ & $2.30$ \\
$9010.965542$ & $-0.56$ & $2.21$ \\
$9010.973732$ & $2.87$ & $2.32$ \\
$9011.918730$ & $-0.84$ & $2.22$ \\
$9012.987404$ & $-7.40$ & $3.00$ \\
$9012.993618$ & $1.11$ & $2.89$ \\
$9014.952306$ & $2.15$ & $2.39$ \\
$9014.959983$ & $8.99$ & $2.31$ \\
$9017.952195$ & $0.52$ & $2.35$ \\
$9017.959774$ & $-3.71$ & $2.43$ \\
$9029.955337$ & $4.42$ & $2.25$ \\
$9029.962978$ & $0.18$ & $2.22$ \\
$9030.952815$ & $-2.23$ & $2.32$ \\
$9030.960649$ & $-0.90$ & $2.30$ \\
$9031.947822$ & $-6.02$ & $2.62$ \\
$9031.955839$ & $-2.77$ & $2.44$ \\
$9032.950886$ & $-4.22$ & $2.26$ \\
$9032.958518$ & $-8.38$ & $2.30$ \\
$9033.958650$ & $-4.60$ & $2.21$ \\
$9033.966256$ & $-2.75$ & $2.13$ \\
$9034.876530$ & $-0.86$ & $2.24$ \\
$9034.884052$ & $-0.05$ & $2.18$ \\
$9035.952939$ & $2.06$ & $2.39$ \\
$9035.960628$ & $0.34$ & $2.31$ \\
$9036.946364$ & $6.51$ & $2.33$ \\
$9036.953908$ & $7.24$ & $2.31$ \\
$9037.954089$ & $4.63$ & $2.53$ \\
$9037.961776$ & $0.57$ & $2.47$ \\
$9052.760439$ & $1.09$ & $2.58$ \\
$9052.767841$ & $-0.41$ & $2.48$ \\
$9054.740438$ & $-6.13$ & $2.58$ \\
$9054.747908$ & $-2.84$ & $2.56$ \\
$9055.737313$ & $-1.26$ & $2.79$ \\
$9055.746024$ & $-0.83$ & $2.84$ \\
$9062.738505$ & $-7.60$ & $2.64$ \\
$9062.746918$ & $-1.06$ & $2.63$ \\
$9064.753007$ & $-6.63$ & $2.57$ \\
$9065.730471$ & $-3.30$ & $2.60$ \\
$9065.738626$ & $1.92$ & $2.60$ \\
$9068.828720$ & $7.37$ & $3.48$ \\
$9068.835744$ & $4.97$ & $3.17$ \\
$9301.100420$ & $-13.15$ & $12.78$ \\
$9301.924685$ & $-4.96$ & $3.52$ \\
$9301.931132$ & $-7.24$ & $4.16$ \\
$9305.012038$ & $3.51$ & $2.55$ \\
$9305.021274$ & $0.05$ & $2.44$ \\
$9321.988167$ & $0.33$ & $3.39$ \\
$9321.991843$ & $6.56$ & $3.47$ \\
$9329.979601$ & $7.24$ & $3.54$ \\
$9329.984625$ & $3.49$ & $3.38$ \\
$9335.998818$ & $1.56$ & $2.38$ \\
$9336.006043$ & $-0.96$ & $2.38$ \\
$9336.903887$ & $1.66$ & $2.31$ \\
$9336.914754$ & $1.94$ & $2.26$ \\
$9352.843703$ & $0.65$ & $2.73$ \\
$9352.850681$ & $1.97$ & $2.70$ \\
$9353.824596$ & $-5.58$ & $2.71$ \\
$9353.829970$ & $-4.19$ & $2.77$ \\
$9354.825976$ & $-0.43$ & $2.59$ \\
$9354.832940$ & $-2.78$ & $2.60$ \\
$9367.811663$ & $1.07$ & $2.90$ \\
$9367.817009$ & $-4.90$ & $2.68$ \\
$9368.804957$ & $-2.86$ & $2.76$ \\
$9368.809560$ & $0.26$ & $2.74$ \\
$9369.812588$ & $2.78$ & $2.80$ \\
$9369.817475$ & $1.25$ & $2.81$ \\
$9370.812156$ & $1.10$ & $2.47$ \\
$9370.818195$ & $7.52$ & $2.55$ \\
$9371.828379$ & $8.15$ & $2.36$ \\
$9371.834659$ & $2.45$ & $2.42$ \\
$9372.819000$ & $2.06$ & $2.48$ \\
$9372.825454$ & $-2.33$ & $2.42$ \\
$9410.740034$ & $0.57$ & $3.07$ \\
$9410.743733$ & $-4.85$ & $3.02$ \\
$9411.743618$ & $-3.48$ & $3.61$ \\
$9411.747263$ & $2.08$ & $3.62$ \\
$9452.777252$ & $8.36$ & $3.56$ \\
$9452.780933$ & $2.07$ & $3.63$ \\
$9453.799319$ & $4.47$ & $3.62$ \\
$9453.803015$ & $9.77$ & $3.61$ \\
$9455.725685$ & $4.62$ & $3.00$ \\
$9455.731175$ & $13.25$ & $3.02$ \\
$9468.728240$ & $9.62$ & $3.12$ \\
$9468.733409$ & $12.50$ & $3.08$ \\
$9486.714323$ & $1.74$ & $3.52$ \\
$9486.719672$ & $4.35$ & $3.13$ \\
\hline
\label{tbl:rv_obs}
\end{longtable}
\twocolumn

\section{Additional Figures}
\begin{figure*}[h]
    \centering
    \includegraphics[scale=0.45]{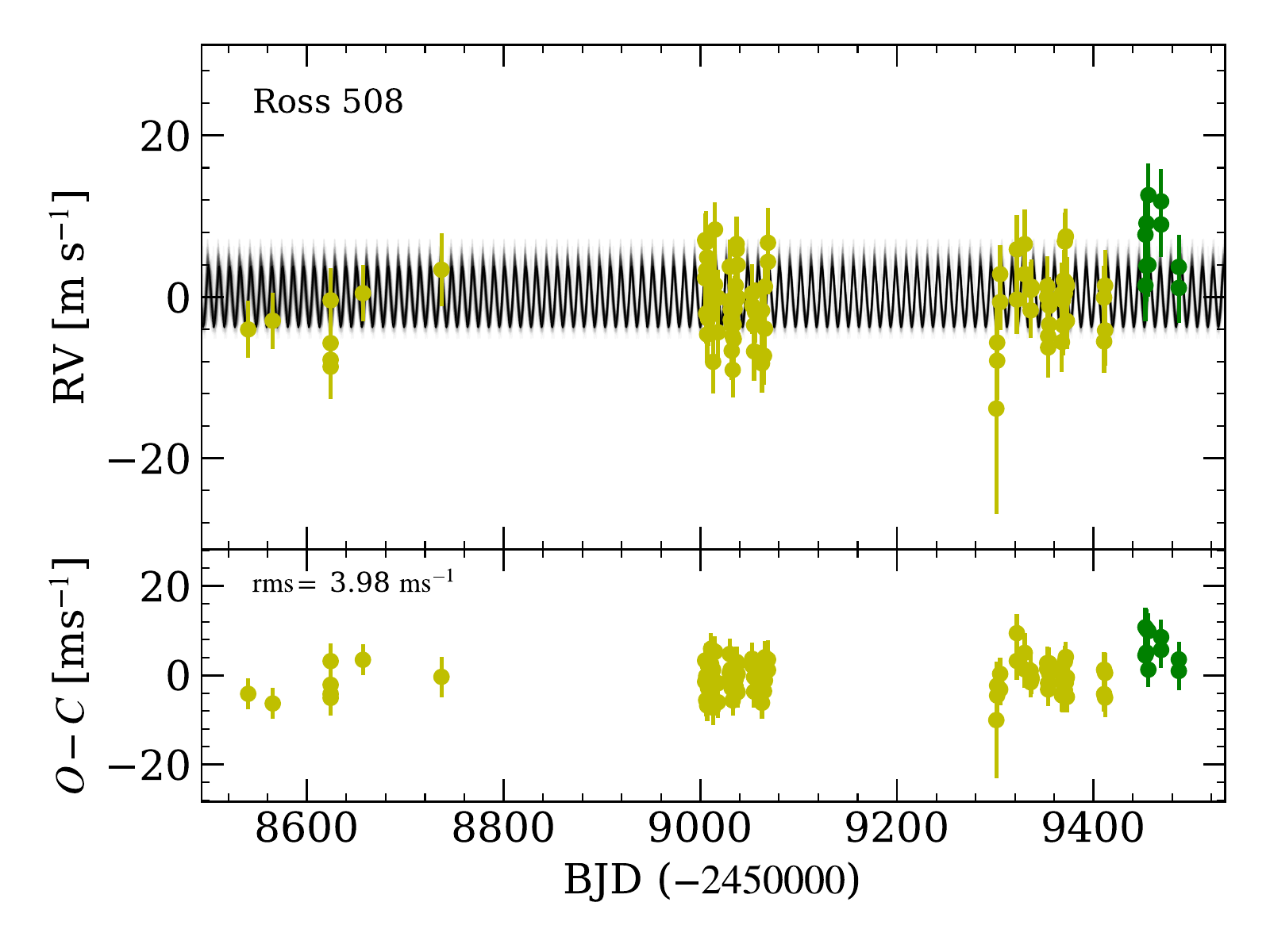}\\
    \includegraphics[scale=0.45]{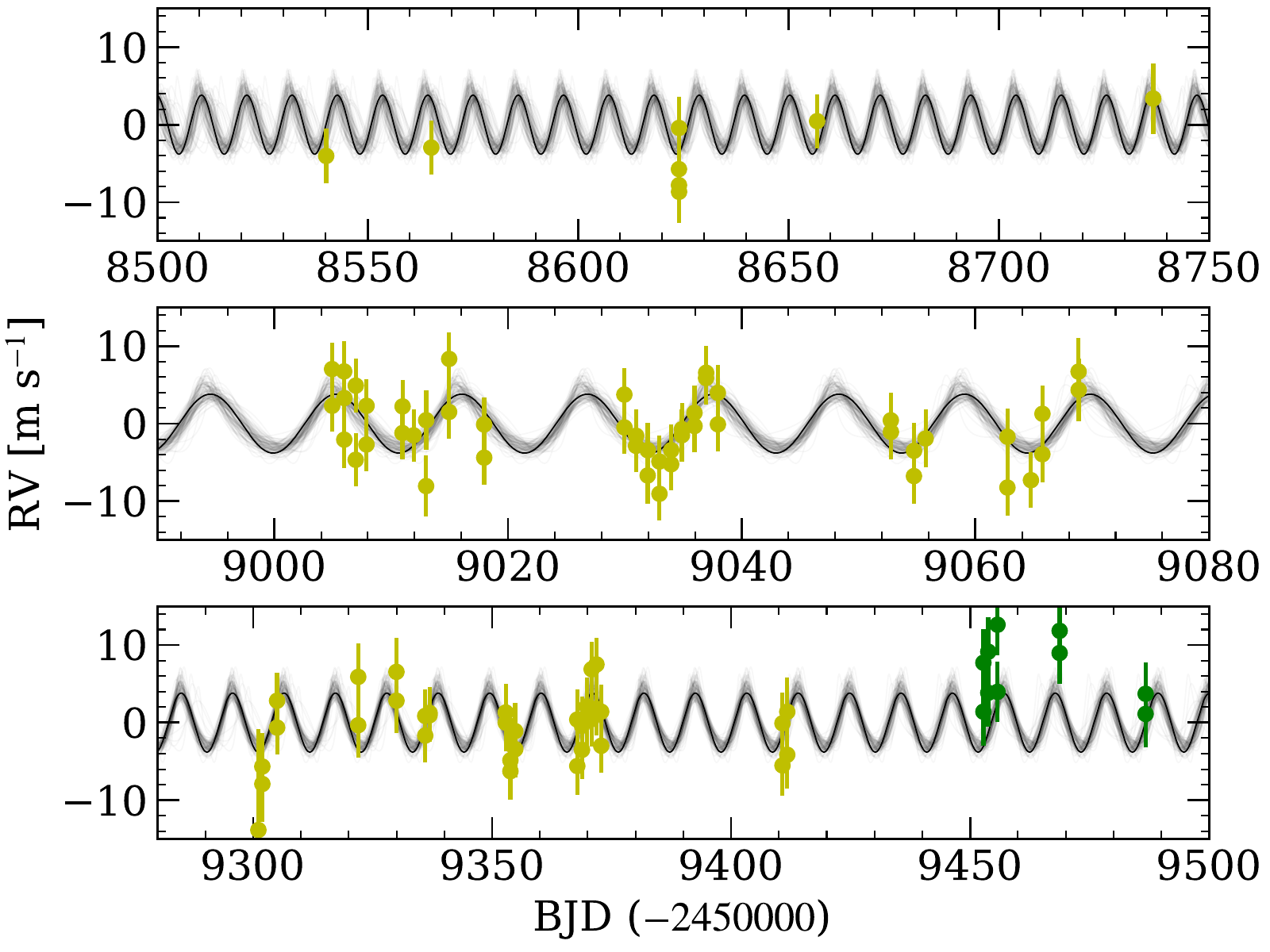}\\
    \includegraphics[scale=0.45]{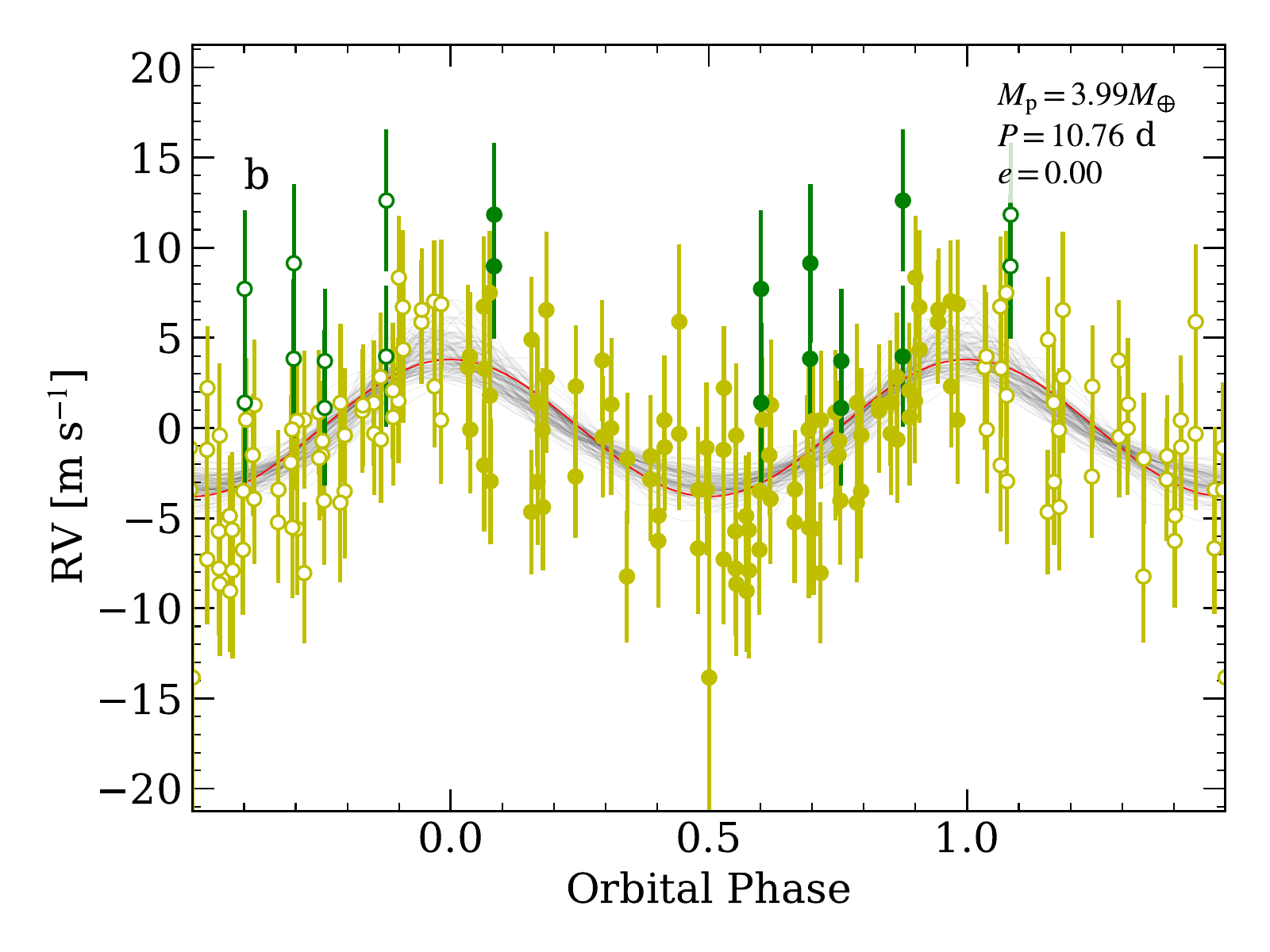}
    \caption{Observed RVs and best-fit RV model of Model A1. See caption to Figure \ref{fig:rv_fit_B1} for details.}
    \label{fig:rv_fit_A1}
\end{figure*}

\begin{figure*}
    \centering
    \includegraphics[scale=0.22]{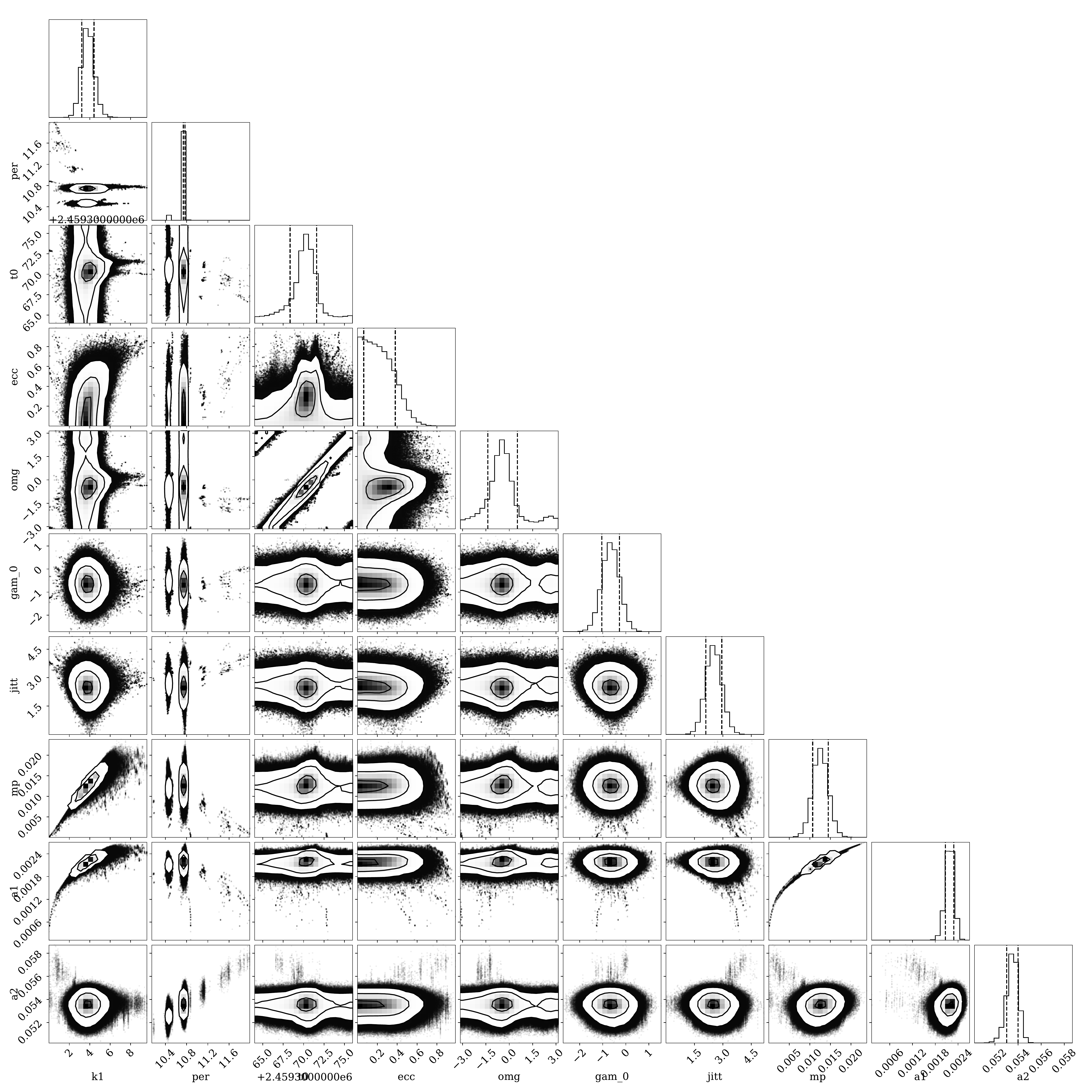}
    \caption{Corner plot of Model A1.}
    \label{fig:corner_A1}
\end{figure*}

\begin{figure*}
    \centering
    \includegraphics[scale=0.45]{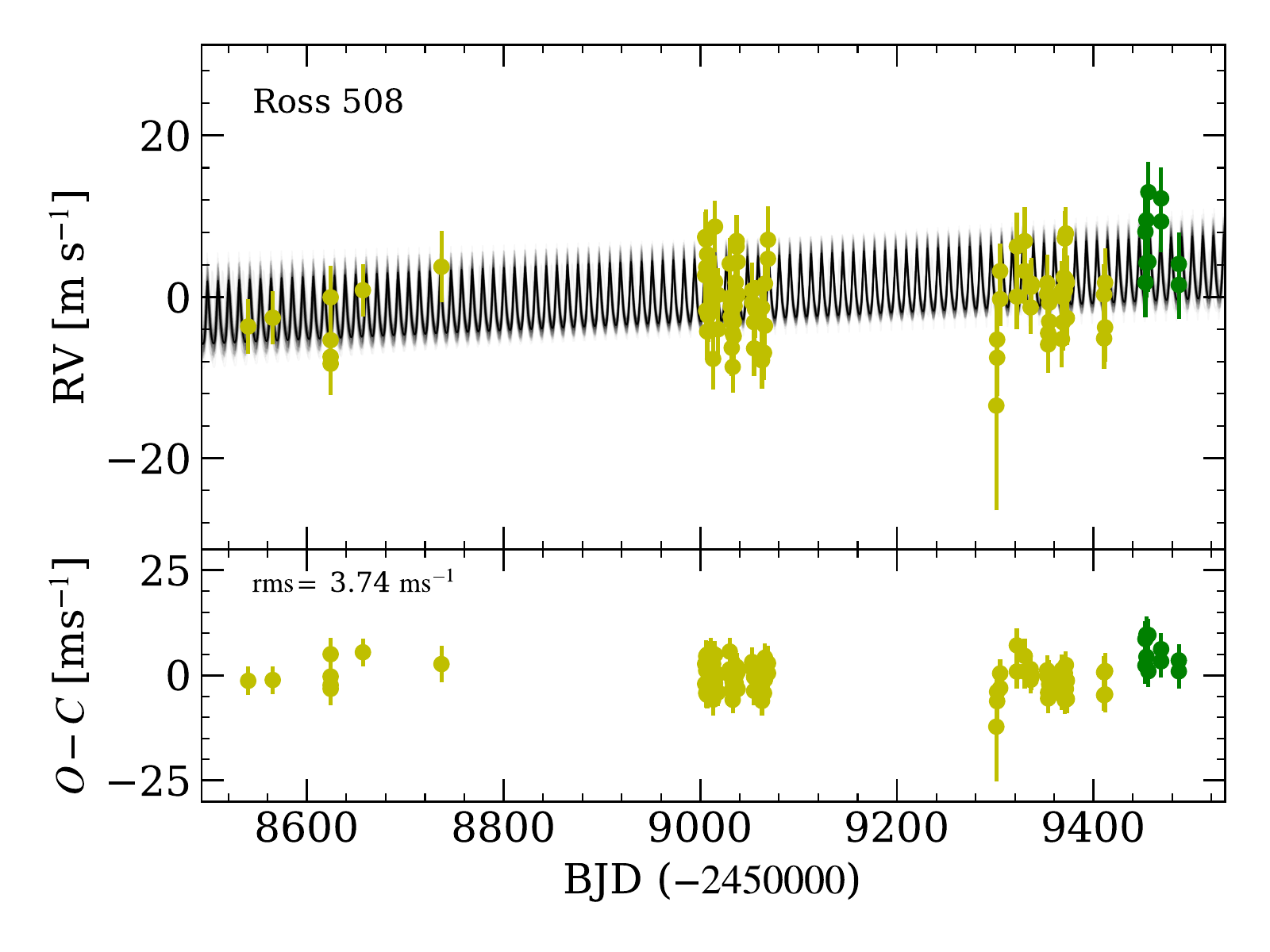}\\
    \includegraphics[scale=0.45]{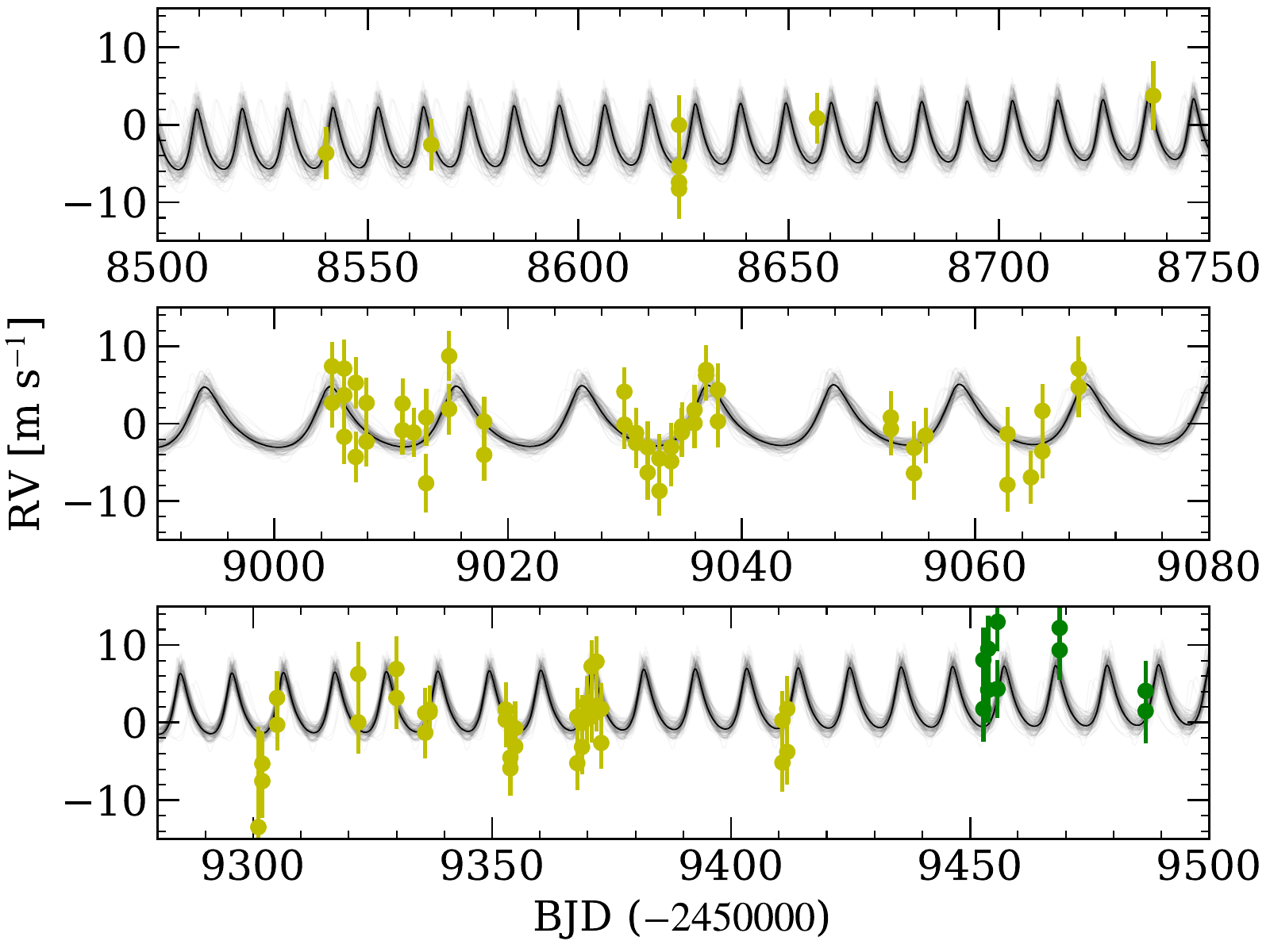}\\
    \includegraphics[scale=0.45]{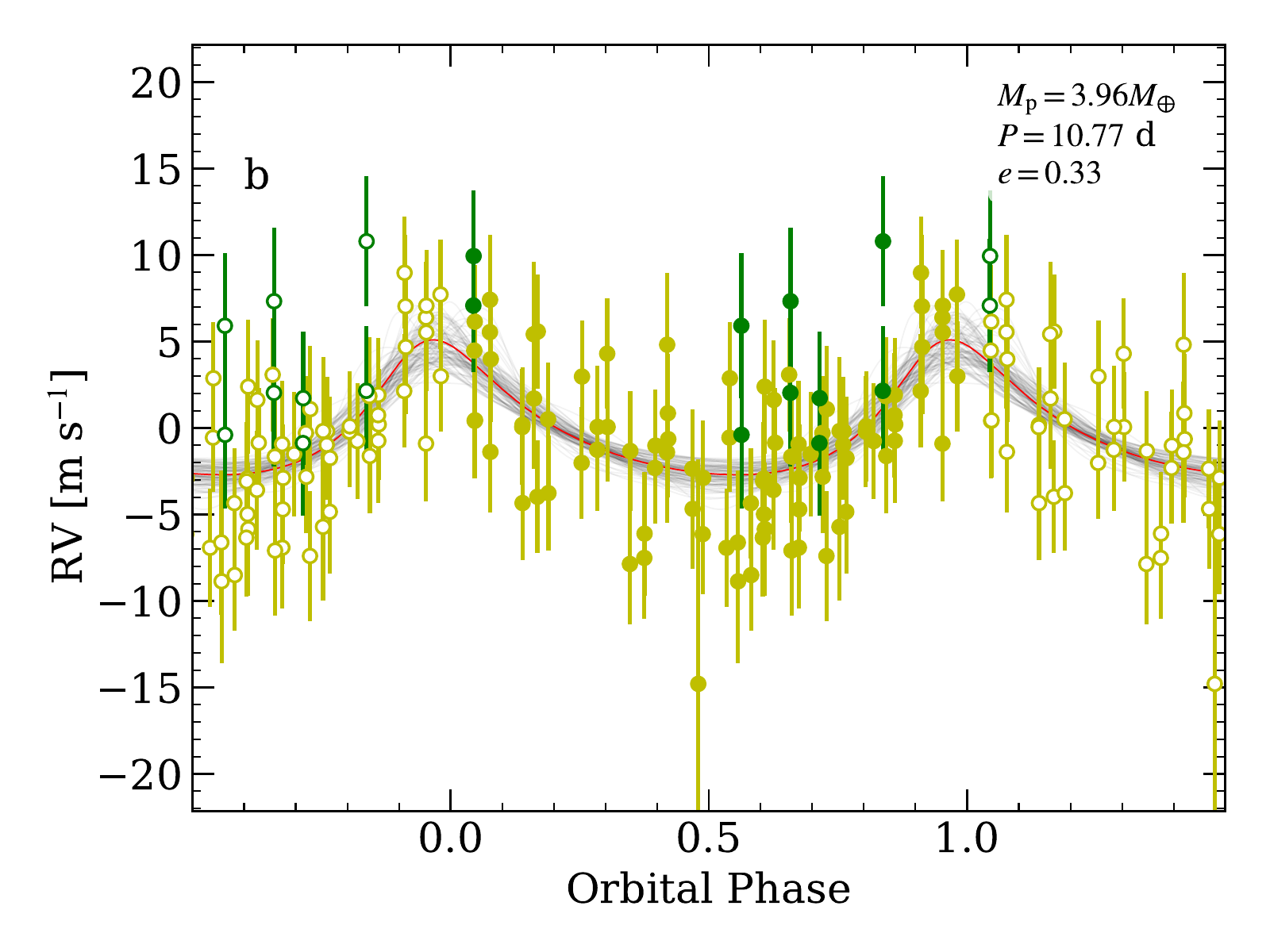}
    \caption{Observed RVs and best-fit RV model of Model A2. See caption to Figure \ref{fig:rv_fit_B1} for details.}
    \label{fig:rv_fit_A2}
\end{figure*}

\begin{figure*}
    \centering
    \includegraphics[scale=0.22]{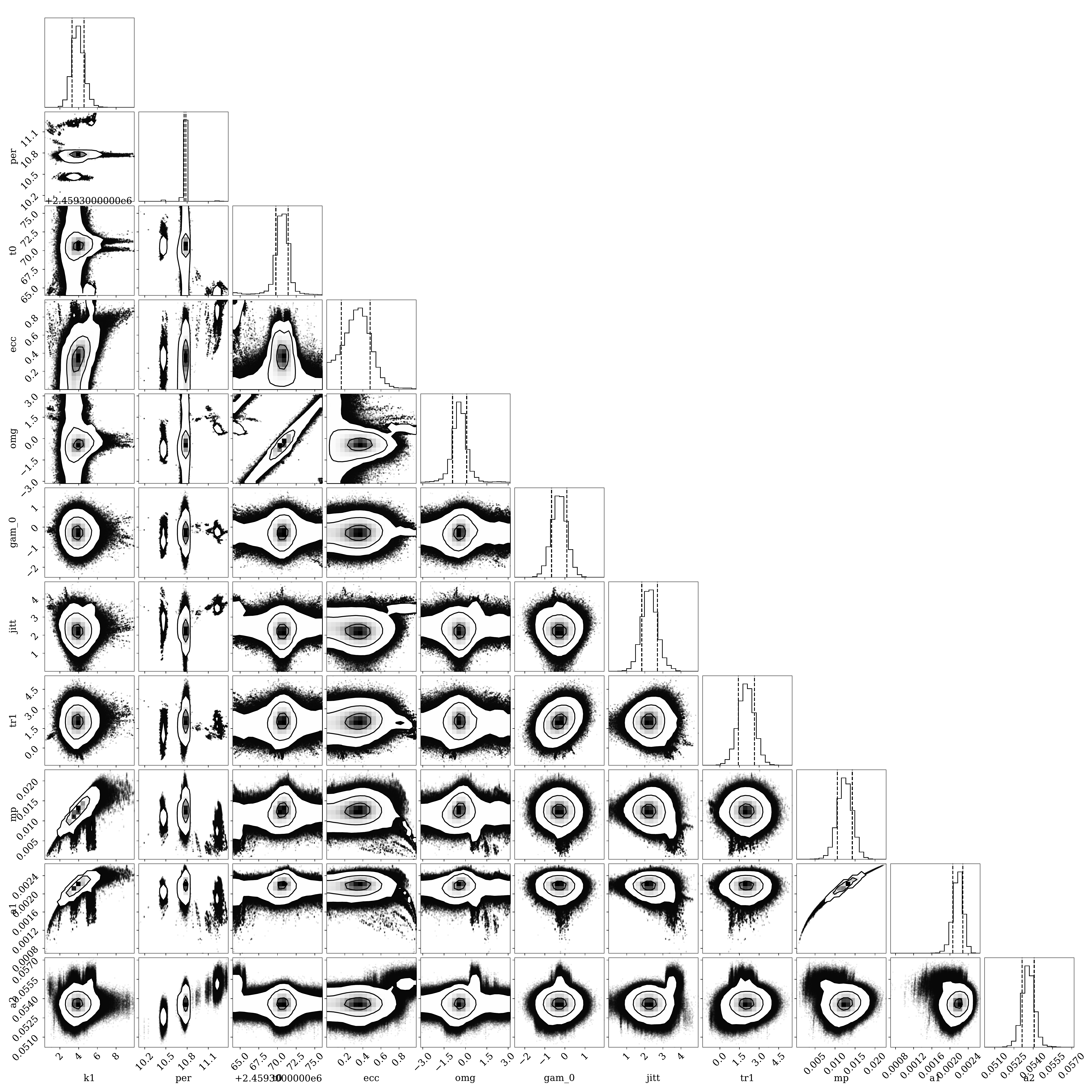}
    \caption{Corner plot of Model A2.}
    \label{fig:corner_A2}
\end{figure*}

\begin{figure*}
    \centering
    \includegraphics[scale=0.45]{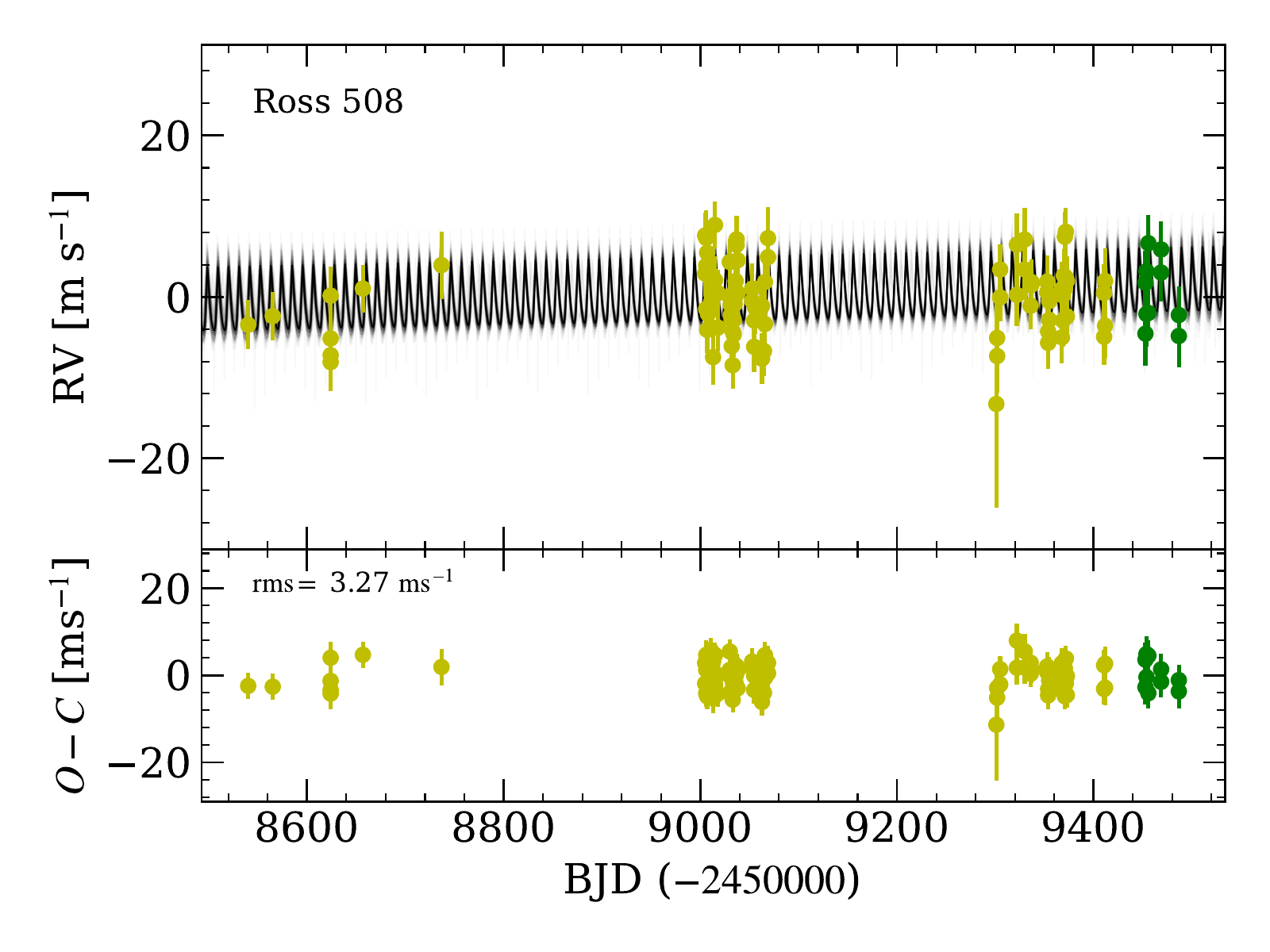}\\
    \includegraphics[scale=0.45]{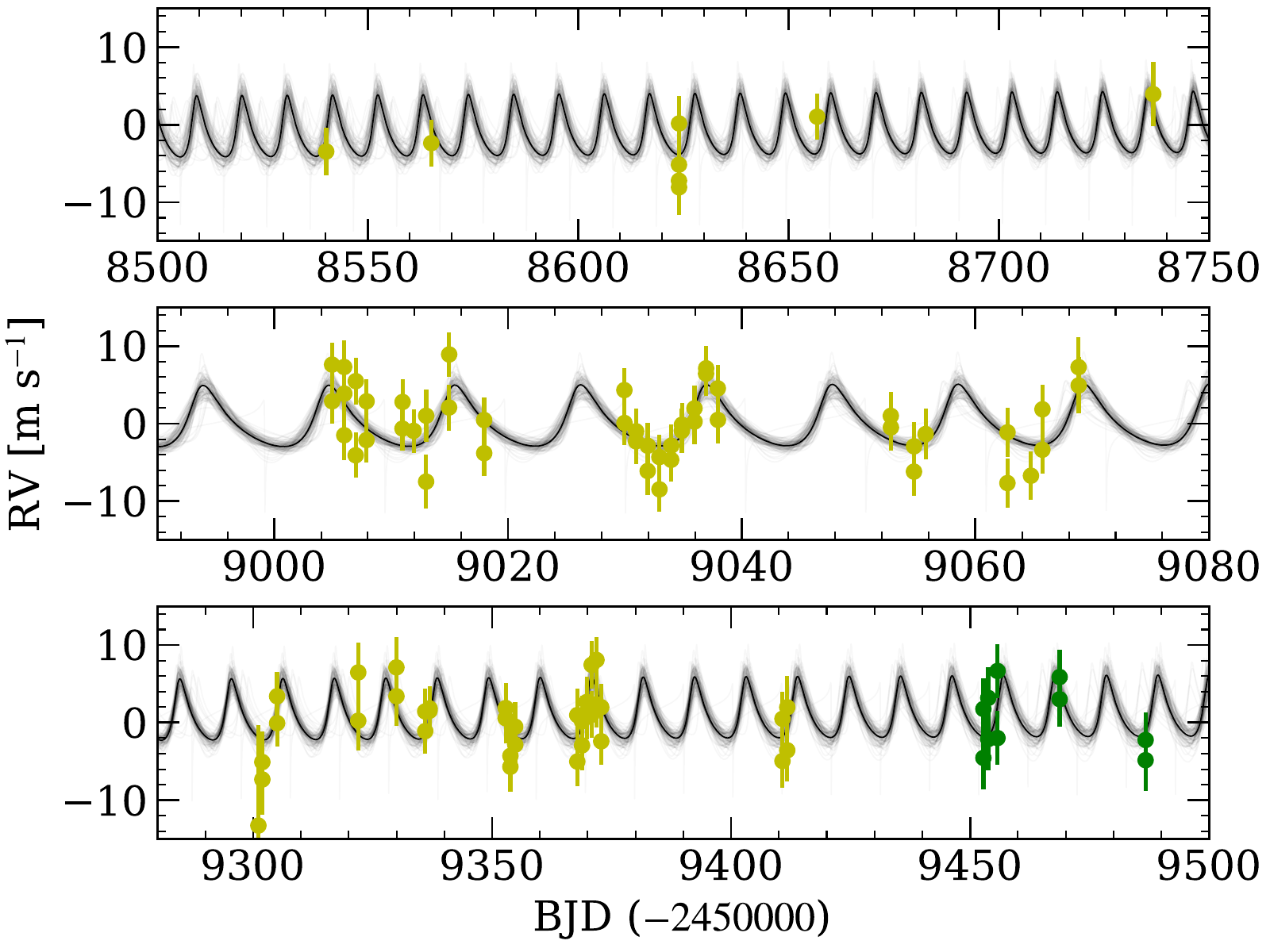}\\
    \includegraphics[scale=0.45]{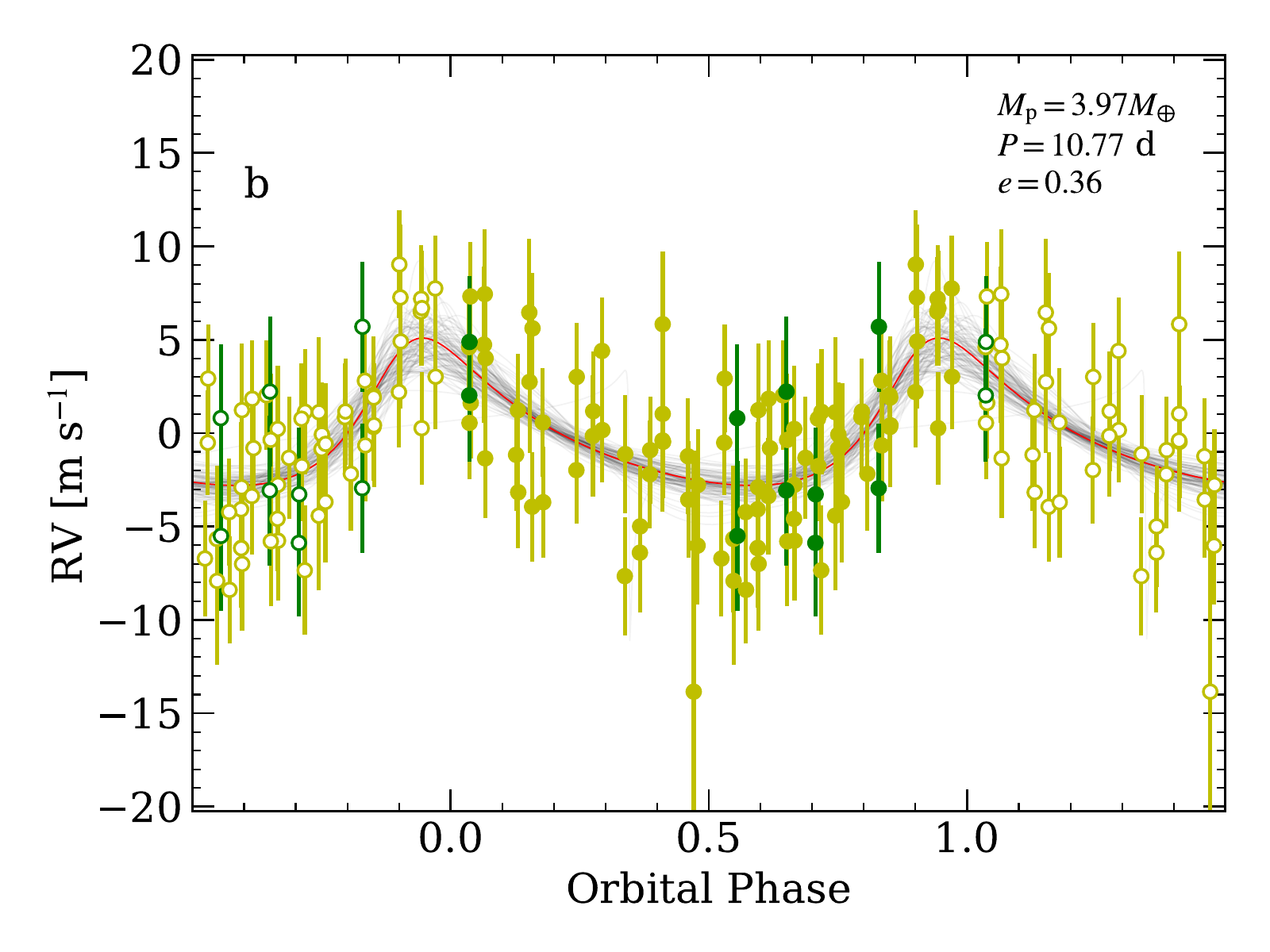}
    \caption{Observed RVs and best-fit RV model of Model B2. See caption to Figure \ref{fig:rv_fit_B1} for details.}
    \label{fig:rv_fit_B2}
\end{figure*}

\begin{figure*}
    \centering
    \includegraphics[scale=0.22]{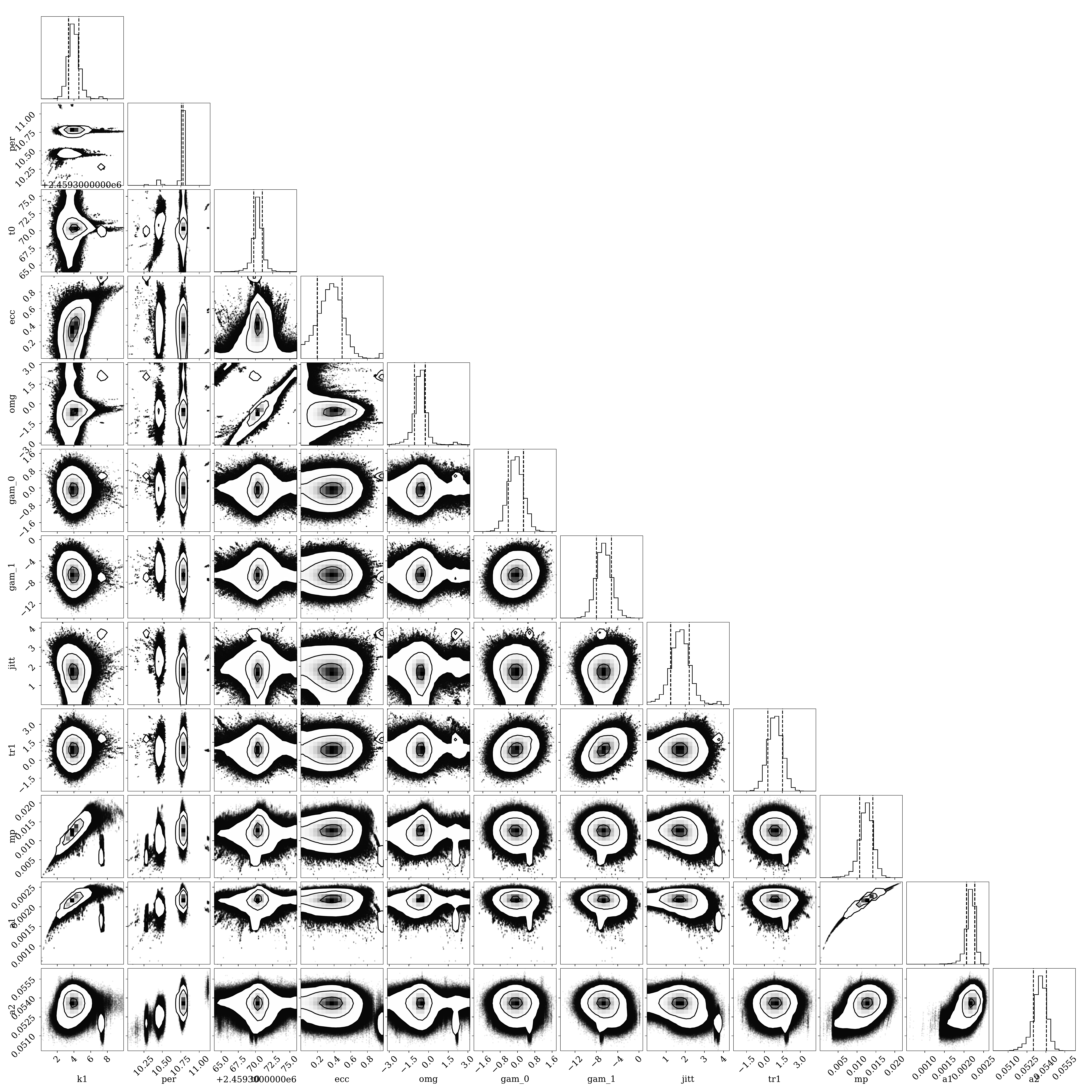}
    \caption{Corner plot of Model B2.}
    \label{fig:corner_B2}
\end{figure*}

Figures \ref{fig:rv_fit_A1} through \ref{fig:corner_B2} show the best-fit orbital solutions and the corner plots. Each pair of figures corresponds to Model A1, A2, and B2, respectively.

\end{document}